\documentclass[aps,prd,showpacs,showkeys,superscriptaddress,twocolumn]{revtex4-1}
\usepackage[colorlinks,linkcolor=blue,anchorcolor=blue,citecolor=blue,urlcolor=blue,breaklinks=true]{hyperref}
\usepackage{graphicx}
\usepackage{amsmath}
\usepackage{amssymb}
\usepackage{slashed}
\usepackage{latexsym}
\usepackage{epsfig}
\usepackage{amsbsy}
\usepackage{array}
\usepackage{changes}
\usepackage{amssymb}
\usepackage{setspace}
\usepackage{bm}
\usepackage{lipsum}
\usepackage{mathrsfs}
\usepackage{float}
\usepackage{color}
\usepackage{subfigure}
\usepackage[T1]{fontenc}
\usepackage{mathptmx}
\DeclareMathAlphabet{\mathcal}{OMS}{cmsy}{m}{n}
\DeclareSymbolFont{largesymbols}{OMX}{cmex}{m}{n}

\begin{document}
\author{Zheng Zhang}
\email{jozhzhang@163.com}
\affiliation{Department of physics, Nanjing University, Nanjing 210093, China}
\author{Chao Shi}
\email{cshi@nuaa.edu.cn}
\affiliation{Department of Nuclear Science and Technology,
Nanjing University of Aeronautics and Astronautics, Nanjing 210016, China}
\author{Xiaofeng Luo}
\email{xfluo@mail.ccnu.edu.cn}
\affiliation{Key Laboratory of Quark and Lepton Physics (MOE) and Institute of Particle Physics, Central China Normal University, Wuhan 430079, China}
\author{Hong-Shi Zong}
\email{zonghs@nju.edu.cn}
\affiliation{Department of physics, Nanjing University, Nanjing 210093, China}
\affiliation{Department of physics, Anhui Normal University, Wuhu, Anhui 241000, China }
\affiliation{Nanjing Proton Source Research and Design Center, Nanjing 210093, China}
\date{\today}

\title{Rotating fermions inside a spherical boundary}

\begin{abstract}
{We apply the canonical quantization procedure to the Dirac field inside a spherical boundary with rotating coordinates. The rotating quantum states with two kinds of boundary conditions, namely, spectral and MIT boundary conditions, are defined. To avoid faster-than-light, we require the speed on the surface to be less than the speed of light. For this situation, the definition of vacuum is unique and identical with the Minkowski vacuum. Finally, we calculate the thermal expectation value of the fermion condensate in a thermal equilibrium rotating fermion field and find it depends on the boundary condition.}
\bigskip


\end{abstract}

\maketitle


\maketitle

\section{Introduction}
The quantization of fields in rotating coordinates attracted some attention since 1980s', which was partly motivated by Hawking's work on black-hole evaporation \cite{Hawking}. Fulling showed that quantization in Rindler coordinates is not identical with that in ordinary Minkowski coordinates and an accelerated observer sees the Minkowski vacuum as a thermal bath \cite{Fulling}. Then Unruh elucidated the relationship between the quantization schemes in Rindler coordinates and the black-hole evaporation, and showed a model particle detector in an accelerated state of motion indeed observes particles in the Minkowski vacuum \cite{Unruh}. It's natural to investigate whether similar effects will occur in other noninertia frames. Letaw and Pfautsch studied the scalar field theory in rotating coordinates \cite{LetPfa1}. They found unlike the uniformly accelerating observer, there is no such fancy effect in a rotating frame. Then Iyer investigated the Dirac field theory in rotating coordinates \cite{Iyer}. He found for Dirac field, quantization scheme in rotating coordinates is inequivalent to the usual Minkowski quantization scheme. However, the systems discussed in \cite{LetPfa1} and \cite{Iyer} are both unbounded, which means the region outside a radius $R$ will have a speed larger than the speed of light. This unphysical property leads to some difficulties of quantization for scalar field \cite{LetPfa1}, and non-unique quantization schemes for Dirac field \cite{rcylinder}. In addition, some problems when calculating the thermal expectations in unbounded rotating systems will occur \cite{Ottewill,Pathology1, rcylinder}. These problems can be solved by constraining the rotating system inside a region whose speed is less than the speed of light. Rotating scalar field bound inside a cylinder surface was investigated in \cite{Ottewill}, and rotating Dirac field bound inside a cylinder surface was investigated in \cite{rcylinder}. It is shown that for rotating systems bound inside a cylindrical boundary, the quantization scheme is identical to that in usual Minkowski coordinates. These results imply that a physically possible cylindrical rotating frame is certainly different from a uniformly accelerating frame where some fancy effects will occur.

To understand the problems occur in quantization of field in rotating coordinates, let us review the usual cannonical quantization procedures in Minkowski coordinates. First, one solves the field equation and finds a set of complete orthogonal modes. These modes are split into positive and negative frequency modes. Then one expands the field operator by the modes and promotes the expansion coefficients into operators. The coefficients of the positive frequency modes are promoted to annihilation operators and the coefficients of the negative frequency modes are promoted to creation operators. Next, one assumes the commutation (anti-commutation) relations of the annihilation operators and the creation operators, and  defines the vacuum as the state annihilated by all the annihilation operators. For a field in Minkowski coordinates, the split of positive and negative modes are clear, the positive (negative) modes have positive (negative) Minkowski energy $E$. However, for field in rotating coordinates, the split of positive modes and negative modes is not as clear as that in Minkowski coordinates. This is because a rotating mode with energy $\widetilde{E}>0$ may have Minkowski energy $E<0$, and vice versa. Should we regard $\widetilde{E}>0$ or $E>0$ modes as positive frequency modes ? For a scalar field, if we want the positive frequency modes have positive Klein-Gordon norm, we must define ${E}>0$ modes as positive frequency modes because the Klein-Gordon norm is proportional to $E$ \cite{Iyer}. But for Dirac field, all modes have positive Dirac norm, so the split of positive and negative modes seems less constrained. If we define $E>0$ modes as positive frequency modes, the vacuum is called nonrotating vacuum, which is identical to the Minkowski vacuum. If we define $\widetilde{E}>0$ modes as positive frequency modes, the vacuum is called rotating vacuum. On unbounded Minkowsi space-time, there exists modes with $E\widetilde{E}<0$, thus making the two vacua inequivalent. If one encloses the field inside the speed of light surface (SOL), one can expect that the modes with $E\widetilde{E}<0$ will not occur, thus the rotating and nonrotating vacuum are equivalent, and the problems are solved. For scalar field, it is shown there are no $E\widetilde{E}<0$ modes when enclosing the field inside a cylindrical boundary with the Dirichlet kind \cite{Vilenkin2}. For Dirac field, one can also prove the same result for spectral and MIT cylindrical boundary conditions \cite{rcylinder}. In this paper, we will prove this result for Dirac field enclosed in a spherical boundary with spectral and MIT kind. It seems there is a general proof for this result, regardless what the shape or kind of the boundary condition is. (Of course, the boundary condition should satisfy some basic requirements, such as keeping the Hamiltonian self-adjoint.) This general proof is not available now, but it's reasonable to believe it exists. 

From the point of view above, the quantization in rotating coordinates is trivial. But it does not mean other aspects of a rotating system are also trivial. For example, macroscopic parity-violating effects can occur in rotating systems \cite{Vilenkin}. When considering the chiral anomaly, there can be chiral vortical effect in noncentral high energy heavy-ion collisions \cite{chiralv2}. To study these effects, one may need thermodynamics and statistical mechanics for rotating systems, whose principles had been introduced by Landau and Lifshitz \cite{Landau5}, and elaborated by Vilenkin \cite{Vilenkin2}. Here we should note that the thermodynamics for rotating systems is not the thermodynamics in noninertial rotating frames. The former deals with what a static observer sees for a rotating system, while the latter deals with things seen by a rotating observer. 
Since the observer is static, one need not use rotating coordinates when discussing about the thermodynamics for rotating systems. But interestingly, problems will occur if the speed of the surface exceeds the speed of light. These problems are also relevant to the modes with $E\widetilde{E}<0$, because the distribution in rotating systems is relevant to $[e^{\beta \widetilde{E}}\pm 1]^{-1}$ \cite{rcylinder}. For unbounded rotating scalar field, the existence of particle modes with $\widetilde{E}=0$ leads to the divergence of the thermal expectations \cite{Pathology1}. For unbounded Dirac field, if one treat $E>0$ modes as particle modes (positive frequency modes), the thermal expectation values will have an unphysical term \cite{Vilenkin2,rcylinder}. These problems can also be cured by enclosing the field inside the SOL. There are some results of thermal expectation values for fermion field bounded by the cylindrical boundary \cite{rcylinder}.

In this paper, we study the Dirac field theory inside a sphere with rotating coordinates. The axis of rotation is selected as $z$-axis. The rigidly-rotating quantum states are constructed. To bound the field inside a sphere, we follow Ref. \cite{rcylinder} to impose two kinds of boundary conditions, the spectral \cite{spectral} and MIT \cite{MIT1} boundary conditions, and give the spectrum in each case. We proved that the rotating and nonrotating vacua are identical when the boundary of the sphere lies within the SOL. The second quantization procedures are performed. The thermal expectation value of fermion condensate is calculated for each boundary condition and possible applications are discussed.

The aim of this paper is of twofold. First, by showing the vacuum of a rotating fermionic field enclosed inside a sphere (whose surface does not exceed the speed of light) is identical to the Minkowski vacuum, together with Ref. \cite{rcylinder} which obtained the same result but for a different geometry boundary, this paper suggests that: any physically possible rotating observer does not see the Minkowski vacuum has strange effects such as Unruh effect. Second, the results of this paper can be used to calculate thermal expectation values for some spherical rotating systems, which may have practical applications. We will discuss the fermion condensate as an example. One advantage of this paper is that it deals with finite rotating systems, which are more close to the real rotating systems, especially when the finite volume has important effects. For example, the results of this paper have potential applications to realistic simulation of the small rotating quark-gluon systems created in heavy-ion collisions \cite{ShuryakQGP,QGPdisco,vorticalfluid,Palhares2011}.

The structure of this paper is organized as follows. In Sec. \ref{II}, we construct the mode solutions of the Dirac equation in unbounded spherical coordinates and review the second quantization procedure. The remainder of the paper considers the bounded space-time with spectral and MIT boundary conditions. In Sec. \ref{Sec3}, we give the mode solutions satisfying the two kind of boundary conditions, show that there is no $E\widetilde{E}<0$ modes if the field is enclosed in SOL, and perform the second quantization procedures. The thermal expectation values of fermion condensate with the two kinds of boundary conditions are calculated in Sec. \ref{IV}. Finally, a summary is included in Sec. \ref{V}. 

\section{Unbounded space-time} \label{II}
In this section, we construct the mode solutions in a rigidly-rotating, unbounded, Minkowski space-time. The Dirac equation is introduced in Sec. \ref{IIA}. The solutions in spherical coordinates are given in Sec \ref{IIB}. In Sec. \ref{IIC}, we discuss the definition of the vacuum.

\subsection{Dirac equation in rotating Minkowski space-time}\label{IIA}
The metric of a rigidly-rotating frame with angular velocity $\Omega$ is given by
\begin{equation}
g_{\mu \nu}=\left(\begin{array}{cccc}
{1-\left(x^{2}+y^{2}\right) \Omega^{2}} & {y \Omega} & {-x \Omega} & {0} \\
{y \Omega} & {-1} & {0} & {0} \\
{-x \Omega} & {0} & {-1} & {0} \\
{0} & {0} & {0} & {-1}
\end{array}\right).
\end{equation}
We adopt the convention that $\hat{i},\hat{j}\cdots=\hat{t},\hat{x},\hat{y},\hat{z}$ and $\mu,\nu\cdots=t,x,y,z$ refer to the Cartesian coordinate in the local rest frame and the general coordinate in the rotating frame, respectively. 

In this paper, we adopt the units $\hbar=c=k_B=1$. The Dirac equation of a fermion with a mass $M$ in the curved spacetime is 
\begin{equation}
[i\gamma^\mu(\partial_\mu+\Gamma^\mu)-M]\psi=0,
\end{equation}
where 
\begin{equation}
\begin{aligned}
&\Gamma_\mu=-\frac{i}{4}\omega_{\mu \hat{i}\hat{j}}\sigma^{\hat{i}\hat{j}},\\
&\omega_{\mu \hat{i}\hat{j}}=g_{\alpha\beta}e^\alpha_{\hat{i}}(\partial_\mu e^{\beta}_{\hat{j}}+\Gamma^{\beta}_{\nu\mu} e^\mu_{\hat{j}}),\\
&\sigma^{\hat{i}\hat{j}}=\frac{i}{2}[\gamma^{\hat{i}},\gamma^{\hat{j}}],
\end{aligned}
\end{equation}
with the Christoffel connection, $\Gamma^\lambda_{\mu\nu}=\frac{1}{2}g^{\lambda\sigma}(g_{\sigma\nu,\mu}+g_{\mu\sigma.\nu}-g_{\mu\nu,\sigma})$, and the gamma matrix in curved space-time, $\gamma^\mu=e^\mu_{\hat{i}}\gamma^{\hat{i}}$. The vierbein $e^\mu_{\hat{i}}$ connects the general coordinate with the Cartesian coordinate in the rest frame, $x^\mu=e^\mu_{\hat{i}}x^{\hat{i}}$. Then the Dirac equation can be reduced to \cite{cylinder}
\begin{equation}
\label{Dirac}
\left[\gamma^{\hat{t}}\left(i \partial_{t}+\Omega J_{z}\right)+i \gamma^{\hat{x}} \partial_{x}+i \gamma^{\hat{y}} \partial_{y}+i \gamma^{\hat{z}} \partial_{z}-M\right] \psi=0.
\end{equation}
where $J_z$ is the z-component of the total angular momentum.
In this paper, we adopt the Pauli-Dirac representation of the gamma matrices:
\begin{equation}\gamma^{\hat{t}}=\left(\begin{array}{cc}
1 & 0 \\
0 & -1
\end{array}\right), \quad \gamma^{\hat{i}}=\left(\begin{array}{cc}
0 & \sigma_{i} \\
-\sigma_{i} & 0
\end{array}\right)\end{equation}
where $\sigma_i$ are Pauli matrices:
\begin{equation}\label{Paulim}
{\sigma_1}=\left(\begin{array}{ll}
{0} & 1 \\
1 &{0}
\end{array}\right),\ \ 
{\sigma_2}=\left(\begin{array}{ll}
{0} & -i \\
i &{\ \ 0}
\end{array}\right),\ \ 
{\sigma_3}=\left(\begin{array}{ll}
{1} & \ \ 0 \\
0 &{-1}
\end{array}\right).
\end{equation}

\subsection{Mode solutions}\label{IIB}
It is observed that the Dirac equation (\ref{Dirac}) is only different from the Dirac equation in Minkowski coordinates by a term about $J_z$. In fact, as we will see, the spherical wave solutions to Eq. (\ref{Dirac}) have the same form with the spherical wave solutions to Dirac equation in Minkowski coordinates. The solutions to Dirac equation in Minkowsi space with respect to spherical coordinates have been reported or partly reported in Refs. \cite{Landau5,Sakurai,Greinerwave,Greiner,kaxinglin}. In this paper, we partly follow Ref. \cite{kaxinglin}.

We assume the form of the solution to Eq. (\ref{Dirac}) as:
\begin{equation}
\psi(x)=u(x)e^{-i\widetilde{E}t}.
\end{equation}
Then we obtain a stationary equation:
\begin{equation}
(-i\gamma^{\hat{0}}\gamma^{\hat{i}}\partial_i+\gamma^{\hat{0}}M-\Omega J_z)u(x)=\widetilde{E}u(x),
\end{equation}
or, by $\boldsymbol{\alpha}=\gamma^{\hat{0}}\boldsymbol{\gamma}, \beta=\gamma^{\hat{0}}$, written as
\begin{equation}\label{eigen}
(-i\mathbf{\alpha}\cdot{\nabla}+\beta M-\Omega J_z)u(x)=\widetilde{E}u(x).
\end{equation}
$\widetilde{E}$ is the total energy in the rotating frame.
We can write the corotating Hamiltonian 
\begin{equation}
\widetilde{H}=i\partial_t=-i\mathbf{\alpha}\cdot{\nabla}+\beta M-\Omega J_z=H-\Omega J_z,
\end{equation}
where $H$ has the same form with the free Hamiltonian in Minkowski coordinates. To solve Eq.(\ref{eigen}), one usually looks for a complete set of commuting operators. The complete set of operators suitable for spherical coordinates is $\{H,J^2,J_z, K\}$. Where $J^2$ is the total angular operator and $K$ is defined by
\begin{equation}
K=\beta(\boldsymbol{L}\cdot \boldsymbol{\Sigma}+1),
\end{equation}
where $\boldsymbol{L}$ is the orbital angular momentum operator and 
\begin{equation}\boldsymbol{\Sigma}=\left(\begin{array}{ll}
\boldsymbol{\sigma} & 0 \\
0 &\boldsymbol{\sigma}
\end{array}\right),
\end{equation}
where $\boldsymbol{\sigma}=(\sigma_1,\sigma_2,\sigma_3)$, which are given in Eqs. (\ref{Paulim}).

One can check operators $H,J^2,J_z, K$ commute with $\widetilde{H}$ and commute with each other. In fact, they are exactly the complete set of commutating operators in Minkowski spherical coordinates \cite{Sakurai}. Thus, the solutions to Eq. (\ref{eigen}) have the same form with the solutions to the Dirac equation in Minkowski coordinates, only the energy $\widetilde{E}$ is different from the Minkowski energy $E$. We label the eigen values of $\{H,J^2,J_z, K\}$ by $\{E,j(j+1),m_j,\kappa\}$, the relation between the corotating energy $\widetilde{E}$ and Minkowski energy $E$ is 
\begin{equation}
\widetilde{E}=E-\Omega m_j.
\end{equation}
The energy difference $\Omega m_j$ can be understood intuitively that a rotating object with angular velocity $\boldsymbol{\Omega}$ and angular momentum $\boldsymbol{J}$ has a rotation energy $\boldsymbol{\Omega}\cdot\boldsymbol{J}$.

For short, we use $k=(E,j,m_j,\kappa)$ to label a eigen state $u_k(r,\theta,\phi)$, which has eigen values $\{E,j(j+1),m_j,\kappa\}$ and corresponds a solution  
\begin{equation}
U_k(t,r,\theta,\phi)=u_k(r,\theta,\phi)e^{-i\widetilde{E}t}
\end{equation}
to Eq. (\ref{Dirac}). 
To solve $u_k(r,\theta,\phi)$, we split it into two parts:
\begin{equation}
u_{k}(r,\theta,\phi)=\left(\begin{array}{c}
u_{k}^{+}(r,\theta,\phi) \\
u_{k}^{-}(r,\theta,\phi)
\end{array}\right).\end{equation}
Use the fact that $u_k(r,\theta,\phi)$ is eigen state of $J^2,J_z,K$, we have
\begin{equation}\label{en2}
\begin{aligned}
&J^2u_k^{\pm}=j(j+1)u_k^{\pm},\\
&(L_z+\frac{1}{2}\sigma_z)u_k^{\pm}=m_j u_k^{\pm},\\
&(\boldsymbol{L}\cdot\boldsymbol{\sigma}+1)u_k^{\pm}=\pm \kappa u_k^{\pm}.
\end{aligned}
\end{equation}
Because $J^2=L^2+\frac{1}{4}\sigma^2+\boldsymbol{L}\cdot \boldsymbol{\sigma}$, $u_k^\pm$ are the eigen  states of $L^2$, and we set the corresponding eigenvalues to be $l^{\pm}(l^{\pm}+1)$:
\begin{equation}
L^2u_k^{\pm}=l^{\pm}(l^{\pm}+1)u_k^{\pm}.
\end{equation}
To solve $u_k^\pm$, we also split into two parts:
\begin{equation}
u_{k}^{\pm}(r,\theta,\phi)=\left(\begin{array}{c}
\phi_{k}^{\pm}(r,\theta,\phi) \\
\varphi_{k}^{\pm}(r,\theta,\phi)
\end{array}\right).\end{equation}
By the second equation in Eqs. (\ref{en2}), we have
\begin{equation}
\left(\begin{array}{cc}
L_{z}+\frac{1}{2}  & 0 \\
0 & L_{z}-\frac{1}{2} 
\end{array}\right)\left(\begin{array}{l}
\phi_k^{\pm} \\
\varphi_k^{\pm}
\end{array}\right)=m_{j} \left(\begin{array}{c}
\phi_k^{\pm} \\
\varphi_k^{\pm}
\end{array}\right).\end{equation}
So we can write $\phi^\pm(r,\theta,\phi)$ and $\varphi^\pm(r,\theta,\phi)$ as 
\begin{equation}\begin{aligned}
&\phi^{\pm}(r,\theta,\phi)=f(r) Y_{l^{\pm}, m_{j}-\frac{1}{2}}(\theta,\phi),\\
&\varphi^{\pm}(r,\theta,\phi)=f^{\prime}(r) Y_{l^{\pm}, m_{j}+\frac{1}{2}}(\theta,\phi).
\end{aligned}\end{equation}
Here we note that we use $f(r)$ ($f'(r)$) as the radial function for both $\phi^\pm$ ($\varphi^\pm$), but it should be different for $\phi^{+}$ ($\varphi^{+}$) and for $\phi^{-}$ ($\varphi^{-}$). Now, let us find the relation between $j,\kappa$ and $l^\pm$. First consider $K^2=J^2+\frac{1}{4}$, we have
\begin{equation}
\kappa=\pm (j+\frac{1}{2}).
\end{equation}
Then by the third equation in Eqs. (\ref{en2}), we have
\begin{widetext}
\begin{equation}\begin{aligned}\label{eq3}
&\left(m_{j}-\frac{1}{2}+1\right) f(r) \mp \kappa f(r)+\sqrt{\left(l^{\pm}+m_{j}+\frac{1}{2}\right)\left(l^{\pm}-m_{j}-\frac{1}{2}+1\right)} f^{\prime}(r)=0,\\
&\sqrt{\left(l^{\pm}-m_{j}+\frac{1}{2}\right)\left(l^{\pm}+m_{j}-\frac{1}{2}+1\right)} f(r)-\left(m_{j}+\frac{1}{2}-1\right) f^{\prime}(r) \mp \kappa f^{\prime}(r)=0.
\end{aligned}\end{equation}
\end{widetext}

Equations (\ref{eq3}) have nonzero solutions when the determinant of coefficients equals to zero, then we get
\begin{equation}
l^{\pm}(l^{\pm}+1)=\kappa(\kappa\mp1).
\end{equation}
That is, when $\kappa=j+\frac{1}{2}>0$,
\begin{equation}\left.\begin{array}{l}
l^{+}=\kappa-1=j-\frac{1}{2} \\
l^{-}=\kappa=j+\frac{1}{2}
\end{array}\right\},\end{equation}
when $\kappa=-(j+\frac{1}{2})<0$,
\begin{equation}\left.\begin{array}{l}
l^{+}=-\kappa=j+\frac{1}{2} \\
l^{-}=-(\kappa+1)=j-\frac{1}{2}
\end{array}\right\}.\end{equation}
There is no solution for $\kappa=0$ since $j$ cannot be negative.
The ratio between $f(r)$ and $f'(r)$ is 
\begin{equation}\frac{f(r)}{f^{\prime}(r)}=-\frac{\sqrt{\left(l^{\pm}+m_{j}+\frac{1}{2}\right)\left(l^{\pm}-m_{j}+\frac{1}{2}\right)}}{m_{j}+\frac{1}{2} \mp \kappa}.\end{equation}
Thus we can write $u_k^\pm$ as follows:\\
when $\kappa>0$,
\begin{equation}
u_k^{+}=f(r)\chi_{j m_j}^{+}, \quad u_k^{-}=g(r) \chi_{j m_j}^{-},
\end{equation}
when $\kappa<0$,
\begin{equation}
u_k^{+}=f(r)\chi_{j m_j}^{-}, \quad u_k^{-}=g(r) \chi_{j m_j}^{+},
\end{equation}
where
\begin{equation}\label{chi}
\begin{aligned}
&\chi_{j m_j}^{+}=\left(\begin{array}{l}
\sqrt{\frac{j+m_j}{2 j}} Y_{j-\frac{1}{2}, m_j-\frac{1}{2}} \\
\sqrt{\frac{j-m_j}{2 j}} Y_{j-\frac{1}{2}, m_j+\frac{1}{2}}
\end{array}\right),\\
&\chi_{j m_j}^{-}=\left(\begin{array}{c}
\sqrt{\frac{j-m_j+1}{2(j+1)}} Y_{j+\frac{1}{2}, m_j-\frac{1}{2}} \\
-\sqrt{\frac{j+m_j+1}{2(j+1)}} Y_{j+\frac{1}{2}, m_j+\frac{1}{2}}
\end{array}\right).
\end{aligned}
\end{equation}
$j=\frac{1}{2},\frac{3}{2},...,\ m_j=-j,-j+1,...,j$.
Now let us look for the radial functions $f(r)$ and $g(r)$. To do this, we need the eigen equation
\begin{equation}\label{en4}
Hu(r,\theta,\phi)=Eu(r,\theta,\phi).
\end{equation}
To solve Eq. (\ref{en4}) in spherical coordinates, we need express $H$ in spherical coordinates. By 
\begin{equation}
\boldsymbol{r}\times(\boldsymbol{r}\times\nabla)=\boldsymbol{r}(\boldsymbol{r}\cdot\nabla)-r^2\nabla=\boldsymbol{r}r\frac{\partial}{\partial r}-r^2\nabla,
\end{equation}
we have 
\begin{equation}
-i\boldsymbol{\alpha}\cdot\nabla=-\frac{i}{r}(\boldsymbol{\alpha}\cdot\boldsymbol{r})\frac{\partial}{\partial r}-\frac{1}{r^2}\boldsymbol{\alpha}\cdot\boldsymbol{r}\times\boldsymbol{L},
\end{equation}
where $\boldsymbol{r}$ is the position vector operator. Then use the identity
\begin{equation}(\boldsymbol{\alpha} \cdot \boldsymbol{A})(\boldsymbol{\Sigma} \cdot \boldsymbol{B})=\gamma^{5} \boldsymbol{A} \cdot \boldsymbol{B}+\mathrm{i} \boldsymbol{\alpha} \cdot \boldsymbol{A} \times \boldsymbol{B}\end{equation}
and the fact $\boldsymbol{r}\cdot \boldsymbol{L}=0$, we have 
\begin{equation}
i\boldsymbol{\alpha}\cdot\boldsymbol{r}\times\boldsymbol{L}=(\boldsymbol{\alpha}\cdot\boldsymbol{r})(\boldsymbol{\Sigma}\cdot\boldsymbol{L})=(\boldsymbol{\alpha}\cdot\boldsymbol{r})(\beta K-1).
\end{equation}
So $H$ can be written as 
\begin{equation}
H=-\frac{i}{r}(\boldsymbol{\alpha}\cdot\boldsymbol{r})[\frac{\partial}{\partial r}-\frac{1}{r}(\beta K-1)]+M\beta.
\end{equation}
Plug it into Eq. (\ref{en4}), and use $\boldsymbol{\sigma}\cdot\frac{\boldsymbol{r}}{r}\chi_{mj}^{\pm}=\chi_{mj}^{\mp}$, we get the equations that the radial functions satisfy:
\begin{equation}\label{en5}
\begin{aligned}
\left(M-E\right) f(r)-\frac{i}{r}(\kappa+1) g(r)-i \frac{\partial g(r)}{\partial r}=0, \\
\left(-M-E\right) g(r)+\frac{i}{r}(\kappa-1) f(r)-i\frac{\partial f(r)}{\partial r}=0,
\end{aligned}
\end{equation}
which can be turned into the form:
\begin{equation}
\begin{aligned}
r^2\frac{\partial^2 f(r)}{\partial r^2}+2r\frac{\partial f(r)}{\partial r}+[p^2 r^2-\kappa(\kappa-1)]f(r)=0,\\
r^2\frac{\partial^2 g(r)}{\partial r^2}+2r\frac{\partial g(r)}{\partial r}+[p^2 r^2-\kappa(\kappa+1)]g(r)=0,
\end{aligned}
\end{equation}
where $p^2=E^2-M^2$. These equations are the spherical Bessel equations, their solutions are:\\
when $\kappa=j+\frac{1}{2}>0$,
\begin{equation}
f(r)=a_k j_{j-\frac{1}{2}}(pr),\ \ \ g(r)=b_k j_{j+\frac{1}{2}}(pr),
\end{equation}
when $\kappa=-(j+\frac{1}{2})<0$,
\begin{equation}
f(r)=a_k j_{j+\frac{1}{2}}(pr),\ \ \ g(r)=b_k j_{j-\frac{1}{2}}(pr).
\end{equation}
Plug the solutions into Eq. (\ref{en5}) and use the following formulas about spherical Bessel functions:
\begin{equation}
j'_n(x)+\frac{n+1}{x}j_n(x)=j_{n-1}(x), \ \ j'_n(x)-\frac{n}{x}j_n(x)=-j_{n+1}(x),
\end{equation}
we get
\begin{equation}
\frac{b_k}{a_k}=\mathrm{sgn}(\kappa)\frac{ip}{E+M}.
\end{equation}
Thus, the solutions are finally written as
\begin{equation}\label{solution}
\begin{aligned}
u_k(t,r,\theta,\phi)=\left[\begin{array}{cc}{\sqrt{\frac{E+M}{2E}}j_{j-\frac{1}{2}}(pr)\chi^+_{jm_j}}\\{i\frac{E}{|E|}}\sqrt{\frac{E-M}{2E}}{j_{j+\frac{1}{2}}(pr)\chi^-_{jm_j}}\end{array}\right], \ \ \ \kappa>0,\\
u_k(t,r,\theta,\phi)=\left[\begin{array}{cc}{\sqrt{\frac{E+M}{2E}}j_{j+\frac{1}{2}}(pr)\chi^-_{jm_j}}\\{-i\frac{E}{|E|}}\sqrt{\frac{E-M}{2E}}{j_{j-\frac{1}{2}}(pr)\chi^+_{jm_j}}\end{array}\right], \ \ \ \kappa<0,
\end{aligned}
\end{equation}
where $\chi_{jm_j}^{\pm}$ are given by Eqs. (\ref{chi}).
The solutions above are not normalized, if we multiply a coefficient $C_k^{\mathrm{free}}=\sqrt{\frac{2}{\pi}}p$, they will be normalized as:
\begin{widetext}
\begin{equation}
\int_0^{\infty}r^2 dr \int_0^{\theta}\sin{\theta} d\theta\int_0^{2\pi}d\phi\  {|C_k^{\mathrm{free}}|}^2 U_k^{\dagger}U_{k'}=\delta(k,k')=\delta_{j,j'}\delta_{m_j,m'_{j'}}\delta_{\kappa,\kappa'}\delta(p-p')\theta(EE').
\end{equation}
\end{widetext}
Anti-particle modes $V_k$ can be obtained from the particle modes through charge conjugation, i.e.:
\begin{equation}
V_k(x)=i\gamma^{\hat{2}} U_k^{*}(x),
\end{equation}
and have the following form:
\begin{equation}
V_k(t,r,\theta,\phi)=v_k(r,\theta,\phi)e^{i\widetilde{E}t},
\end{equation}
where 
\begin{equation}\label{anti}
v_k(r,\theta,\phi)=v_{Ejm_j\kappa}(r,\theta,\phi)=(-1)^{m_j+\frac{1}{2}}\frac{iE}{|E|}u_{\overline{k}}(r,\theta,\phi),
\end{equation}
where 
\begin{equation}
\overline{k}=(-E,j,-m_j,-\kappa).
\end{equation}

\subsection{Second quantization}\label{IIC}
As we discussed in the Introduction, the vacuum of Dirac field in rotating coordinates is not uniquely defined, which comes from the freedom to choose the "particle" and "anti-particle" modes. For nonrotating vacuum, particle modes have Minkowski energy $E>0$. For rotating vacuum, particle modes have corotating energy $\widetilde{E}>0$. The difference between rotating vacuum and nonrotating vacuum arises from the modes with $E\widetilde{E}<0$. By enclosing the system inside the SOL, the modes with $E\widetilde{E}<0$ can be eliminated, which has been proved with cylindrical boundary \cite{rcylinder}. We will also prove it for spherical boundary in this paper. Thus, the rotating vacuum and the nonrotating vacuum are equivalent.

Assuming that there is no modes with $E\widetilde{E}<0$, second quantization can be performed by expanding the field operator $\psi(x)$ as:
\begin{equation}\psi(x)=\sum_{k} \theta\left(E_{k}\right)\left[U_{k}(x) \mathbf{b}_{k}+V_{k}(x) \mathbf{d}_{k}^{\dagger}\right],\end{equation}
where $\theta(E_k)$ is the step function which ensures the Minkowski energy $E_k$ is positive and 
\begin{equation}
\sum_k =\sum_{j=1/2}^{\infty}\ \ \sum_{m_j=-j}^{j}\ \ \sum_{\kappa=\pm(j+1/2)}\ \ \int_{|E|>M} dE.
\end{equation}
The one-particle operators $\mathbf{b}_{k}$ and $\mathbf{d}_{k}^{\dagger}$ obey canonical anti-commutation relations:
\begin{equation}
\{\mathbf{b}_k,\mathbf{b}_{k'}^\dagger\}=\delta(k,k'),\ \ \ \{\mathbf{d}_k,\mathbf{d}_{k'}^\dagger\}=\delta(k,k'). 
\end{equation}
All other anti-commutation relations are zero. The vacuum state $|0\rangle$ is defined by 
\begin{equation}
\mathbf{b}_k |0\rangle=\mathbf{d}_k|0\rangle=0.
\end{equation}
In the next section, we will investigate the Dirac field enclosed by two kinds of boundary conditions, namely, spectral and MIT boundary conditions.

\section{Bounded space-time}\label{Sec3}
This paper focus on the quantum fermion field with rotating coordinates, inside a sphere which has radius $R$. To avoid the exceeding of speed of light, we require $R\Omega<1$.

To enclose the field inside the sphere, we consider two kinds of boundary conditions: the spectral and MIT boundary conditions. In Sec. \ref{IIIA} and Sec. \ref{IIIB}, the spectral and MIT boundary conditions are introduced respectively. For each case, the spectrum is derived and the vacuum state is discussed. And for each case, we show the rotating vacuum and nonrotating vacuum coincide.

\subsection{Spectral boundary conditions}\label{IIIA}
Before introducing the spectral boundary condition, we first discuss the constraint on the behavior of the field on the boundary due to the requirement of the self-adjointness of the Hamiltonian. Here we follow the discussion in \cite{rcylinder}.

The Hamiltonian is a self-adjoint operator, that is :
\begin{equation}\label{self}
\langle\psi,\widetilde{H}\chi\rangle=\langle \widetilde{H}\psi,\chi\rangle.
\end{equation}
Since $\widetilde{H}=i\partial_t$, Eq. (\ref{self}) is equivalent to 
\begin{equation}\label{selff}
\partial_t\langle\psi,\chi\rangle=0.
\end{equation}
In the special case $\chi=\psi$, Eq. (\ref{selff}) implies the conservation of particle numbers.
In the rotating frame, one has \cite{rcylinder}
\begin{equation} \label{self0}
\partial_{t}\langle\psi, \chi\rangle=-\int_{\partial V} d \Sigma_{i} \sqrt{-g} \bar{\psi} \gamma^{\hat{i}} \chi,\end{equation}
where $\partial V$ is the 2 dimensional boundary of the 3 dimensional volume $V$. Thus in the spherical coordinates, the self-adjointness of the Hamiltonian requires:
\begin{equation}\label{self2}
R^2 \int_0^{\pi} \sin{\theta}d\theta \int_0^{2\pi} d\phi \ \bar{\psi} \gamma^{\hat{r}} \chi|_{r=R}=0,
\end{equation}
where  $\gamma^{\hat{r}}=\gamma^{\hat{1}}\sin{\theta}\cos{\phi}+\gamma^{\hat{2}}\sin{\theta}\sin{\phi}+\gamma^{\hat{3}}\cos{\theta}$.

To implement the spectral boundary condition, we write the solution $\psi$ of the Dirac equation with $z$-angular momentum $m$ as:
\begin{equation}
\psi(x)=e^{i\phi m}\times (e^{-\frac{i}{2}\phi} \psi^1_m \ \ e^{\frac{i}{2}\phi} \psi^2_m\ \ e^{-\frac{i}{2}\phi} \psi^3_m\ \  e^{\frac{i}{2}\phi} \psi^4_m)^{T}.
\end{equation}
The inner product of $\psi$ and another solution $\chi$ with $z$-angular momentum $m'$ is time-invariant if:
\begin{equation}\label{sp1}
\begin{aligned}
\int_0^{\pi}\ \sin{\theta}d\theta (\psi^{4*}_m(\mathrm{s}\chi^1_m-\mathrm{c}\chi^2_m)+\psi^{3*}_m(\mathrm{s}\chi^2_m+\mathrm{c}\chi^1_m)\\+\psi^{2*}_m(\mathrm{s}\chi^3_m-\mathrm{c}\chi^4_m)+\psi^{1*}_m(\mathrm{s}\chi^4_m+\mathrm{c}\chi^3_m))\delta_{mm'}=0,
\end{aligned}
\end{equation}
where $\mathrm{c},\mathrm{s}$ is the short writing of $\mathrm{cos}\theta, \mathrm{sin}\theta$. The inner product of the charge conjugate $\psi_c=i\gamma^{\hat{2}}\psi^{*}$ of $\psi$ and $\chi$ must also be time-invariant. That is:
\begin{equation}\label{sp2}
\begin{aligned}
\int_0^{\pi}\ \sin{\theta}d\theta (\psi^{1}_{-m}(\mathrm{s}\chi^1_m-\mathrm{c}\chi^2_m)-\psi^{2}_{-m}(\mathrm{s}\chi^2_m+\mathrm{c}\chi^1_m)\\-\psi^{3}_{-m}(\mathrm{s}\chi^3_m-\mathrm{c}\chi^4_m)+\psi^{4}_{-m}(\mathrm{s}\chi^4_m+\mathrm{c}\chi^3_m))\delta_{m,-m'}=0.
\end{aligned}
\end{equation}
To satisfy both Eq. (\ref{sp1}) and Eq. (\ref{sp2}), we can set
\begin{equation}\label{sp3}
\begin{array}{ll}
\psi_{m}^{3}|_{r=R}=\psi_{m}^{4}|_{r=R}=0, & \ \  m>0 ,\\
\psi_{m}^{1}|_{r=R}=\psi_{m}^{2}|_{r=R}=0, & \ \  m<0,
\end{array}\end{equation}
which we call spectral boundary condition. One can also set the third and fourth components zero when $m<0$ , and the first and second components zero when $m>0$. Here we only discuss the implementation in Eq. (\ref{sp3}). We expect the other implementation gives similar results for expectation values.

\subsubsection{Discretization of the momentum}
Apply Eq. (\ref{sp3}) to the mode solutions  (\ref{solution}) requires the momentum $p$ must be discretized  by:
\begin{equation}p_{jm\kappa,i} R=\left\{\begin{array}{ll}
\xi_{j+\frac{1}{2},i} & m\kappa>0 \\
\xi_{j-\frac{1}{2},i} & m\kappa<0
\end{array}\right.,\end{equation}
where $\xi_{n,i}$ is the $i$th nonzero root of the spherical Bessel function $j_n(x)$. Thus, the mode solutions of Dirac equation with spectral boundary condition can be written as:
\begin{equation}\label{spmode}
U_k^{\mathrm{sp}}(x)=C_k^{\mathrm{sp}}U_k(x),
\end{equation}
where $k$ includes a new index $i$:
\begin{equation}\label{spk}
k=(E,j,m_j,\kappa,i),
\end{equation}
and $E=\pm\sqrt{p^2+M^2}$ is the Minkowski energy. The constants $C_k^{\mathrm{sp}}$ will be calculated in Sec. \ref{IIIA3} to make the modes have unit norm. 

\subsubsection{Energy spectrum}

Here we will show there is no $E\widetilde{E}<0$ modes in the particle spectrum, then the rotating and nonrotating vacua are equivalent. But before the proof, let us explain the physical meaning of this conclusion briefly.

In the picture of the Dirac sea, the Minkowski vacuum is the state with all negative modes $E<0$ occupied. But the mode with Minkowski energy $E$ has rotating energy $\widetilde{E}$ as a rotating observer sees. If $\widetilde{E}>0$, this occupied mode will be identified as a particle by the rotating observer. That is, the rotating observer will see the Minkowski vacuum contains particles. However, if the condition $E\widetilde{E}>0$ is satisfied for all modes, the rotating observer will see nothing in the Minkowski vacuum. 

To demonstrate there is no $E\widetilde{E}<0$ modes for the spectral boundary condition, we use the property of the first zero of the spherical Bessel function \cite{mathBessel1}:
\begin{equation}
\xi_{n,1}>n+1.
\end{equation}
Thus for $E>0$, we have:
\begin{equation}
ER \geq pR>\xi_{j-\frac{1}{2},1}>j+\frac{1}{2}>m_j.
\end{equation}
So 
\begin{equation}\label{pr}
\widetilde{E}R=ER-\Omega m_jR > (1-\Omega R) m_j.
\end{equation}
If $\Omega R\leq 1$, then $E\widetilde{E}>0$ for all $M,j,m_j,\kappa,i$. Similarly, when $E<0$, we can also verify that $E\widetilde{E}>0$ for all $M,j,m_j,\kappa,i$. Thus, the rotating and nonrotating vacua are equivalent. Here we can see the key point in the proof is that, by enclosing the system inside the sphere, the momentum is discretized and thus has a nonzero minimum value.

Since $E\widetilde{E}<0$ modes do not appear, we can perform the second quantization procedure as introduced in Sec. \ref{IIC}.
\subsubsection{Normalization}\label{IIIA3}

Before performing the second quantization, we have to calculate the normalization constant $C_k^{\mathrm{sp}}$. The inner product for two particle modes $U_k^{\mathrm{sp}}$ and $U_{k'}^{\mathrm{sp}}$ is:
\begin{equation}\label{ort1}
\begin{aligned}
\langle U_k^{\mathrm{sp}},U_{k'}^{\mathrm{sp}}\rangle &= C_k^{\mathrm{sp}*}C_{k'}^{\mathrm{sp}} \delta(k,k')\mathfrak{I}_{j+\frac{1}{2}}^{+},
\end{aligned}
\end{equation}
where
\begin{equation}
\delta(k,k')=\delta_{jj'}\delta_{m_j,m_j'}\delta_{\kappa,\kappa'}\delta_{ii'}\theta(EE'),
\end{equation}
and
\begin{widetext}
\begin{equation}\begin{aligned}\label{inte}
&\mathfrak{I}_{n+1}^{+}=\int_{0}^{R} d r r^2 \frac{1}{2}\left[j_{n}^{2}(p r)+j_{n+1}^{2}(p r)\right]=\frac{R^{3}}{2}\left[j_{n+1}^{2}(p R)-\frac{2 (n+1)}{p R} j_{n}(p R) j_{n+1}(p R)+j_{n}^{2}(p R)\right], \\
&\mathfrak{I}_{n+1}^{-}=\int_{0}^{R} d r r^2 \frac{1}{2}\left[j_{n}^{2}(p r)-j_{n+1}^{2}(p r)\right]=\frac{R^2}{2 p} j_{n}(p R) j_{n+1}(p R).
\end{aligned}\end{equation}
\end{widetext}
When $m\kappa>0$, $\mathfrak{I}_{j+\frac{1}{2}}^{+}=\frac{R^3}{2}j^2_{j-\frac{1}{2}}(\xi_{j+\frac{1}{2},i})$,
we take
\begin{equation}
C_{Ejm\kappa i}^{\mathrm{sp}}=\frac{\sqrt{2}}{\sqrt{R^3}|j_{j-\frac{1}{2}}(\xi_{j+\frac{1}{2},i})|},\ \ \ m\kappa>0.
\end{equation}
When $m\kappa<0$, $\mathfrak{I}_{j+\frac{1}{2}}^{+}=\frac{R^3}{2}j^2_{j+\frac{1}{2}}(\xi_{j-\frac{1}{2},i})$,
we take
\begin{equation}
C_{Ejm\kappa i}^{\mathrm{sp}}=\frac{\sqrt{2}}{\sqrt{R^3}|j_{j+\frac{1}{2}}(\xi_{j-\frac{1}{2},i})|},\ \ \ m\kappa<0.
\end{equation}
The anti-particle modes are related to particle modes by Eq. (\ref{anti}):
\begin{equation}
V_k^{\mathrm{sp}}(t,r,\theta,\phi)=(-1)^{m_j+\frac{1}{2}}\frac{iE}{|E|}U_{\overline{k}}^{\mathrm{sp}}(t,r,\theta,\phi),
\end{equation}
where 
\begin{equation}
\overline{k}=(-E,j,-m,-\kappa,i).
\end{equation}
Since the particle modes are normalized (the above calculation is valid for both $E>0$ and $E<0$), so are the anti-particle modes. One can check $U_k^{\mathrm{sp}}$ has the same normalization constant with its charge conjugate $V_k^{\mathrm{sp}}$.

\subsubsection{Second quantization}

To perform the second quantization procedures, we first expand the field in terms of the normalized modes:
\begin{equation}\psi_{\mathrm{sp}}=\sum_{k} \theta\left(E\right)\left[U_{k}^{\mathrm{sp}} \mathbf{b}_{k}^{\mathrm{sp}}+V_{k}^{\mathrm{sp}} \mathbf{d}_{k}^{\mathrm{sp} \dagger}\right],\end{equation}
where $k$ is defined in Eq. (\ref{spk}) and 
\begin{equation}\sum_{k} = \sum_{j=1/2}^{\infty}\  \sum_{m_{j}=-j}^{j}\ \  \sum_{\kappa=\pm({j}+ 1 / 2)}\ \sum_{i=1}^{\infty} \sum_{E=\pm\left|E\right|}.\end{equation}
The vacuum $|0^{\mathrm{sp}}\rangle$ for the spectral boundary condition is defined by
\begin{equation}
\mathbf{b}_k^{\mathrm{sp}}|0^{\mathrm{sp}}\rangle=\mathbf{d}_k^{\mathrm{sp}}|0^{\mathrm{sp}}\rangle=0.
\end{equation}

\subsection{MIT boundary conditions}\label{IIIB}
The MIT boundary condition was firstly introduced in \cite{MIT1}. It satisfies Eq. (\ref{self2}) by setting 
\begin{equation}\label{cg}
i\slashed{n}\psi(x_b)=\varsigma \psi(x_b),
\end{equation}
where $n_\mu$ is the normal to the boundary and $\slashed{n}=\gamma^{\mu}n_\mu$. The coefficient $\varsigma$ can take the general form \cite{Lutken}:
\begin{equation}
\varsigma=\exp \left(-i \gamma_{5} \Theta\right)=\cos \Theta-i \gamma_{5} \sin \Theta,
\end{equation}
where $\Theta$ is the chiral angle. Here we only consider the cases when $\Theta=0$ (MIT) and $\Theta=\pi$ (chiral), that is:
\begin{equation}\varsigma=\left\{\begin{array}{l}
1  \ \ \ \ \  (\mathrm{MIT})\\
-1 \ \ \ (\mathrm{chiral})\ .
\end{array}\right.\end{equation}

One can check that the MIT boundary condition Eq.~(\ref{cg}) makes the normal component of the fermionic current $j^\mu=\overline{\psi}\gamma^\mu\psi$ to be zero on the surface, i.e.,
\begin{equation}
n_\mu j^\mu(x_b)=0
\end{equation}

In spherical coordinates, the boundary condition (\ref{cg}) can be written as:
\begin{equation}\label{MIT}
-i\gamma^{\hat{r}}\psi(x_b)=\varsigma \psi(x_b).
\end{equation}
One can check if $\psi(x)$ satisfies this boundary condition, so does its charge conjugation $i\gamma^{\hat{2}}\psi^{*}(x)$.

\subsubsection{Discretization of the momentum}
Substitute the solutions Eqs. (\ref{solution}) into the boundary condition Eq. (\ref{MIT}), and use the identity $(\boldsymbol{\sigma}\cdot\hat{r})\chi_{mj}^{\pm}=\boldsymbol{\sigma}\cdot\frac{\boldsymbol{r}}{r}\chi_{mj}^{\pm}=\chi_{mj}^{\mp}$, we can get the equation for the allowed momentum:
\begin{equation}\label{eigenp}
j_{l_\kappa}(pR)=\operatorname{sgn}(\kappa)\frac{\varsigma p}{E+M}j_{\overline{l}_\kappa}(pR),
\end{equation}
where
\begin{equation}
\begin{aligned}
&l_{\kappa}=\left\{\begin{array}{cl}{\kappa-1}& {\text { for } \kappa>0} \\ {-\kappa} & {\text { for } \kappa<0}\end{array}\right.,\\
&\overline{l}_{\kappa}=\left\{\begin{array}{cc}{\kappa} & {\text { for } \kappa>0} \\ {-\kappa-1} & {\text { for } \kappa<0}\end{array}\right..
\end{aligned}
\end{equation}
Here we note our equation (\ref{eigenp}) is different from that in \cite{Greiner} by a minus sign, because our definition of $\kappa$ is different from theirs by a minus sign. We label the $i$th nonzero root of Eq. (\ref{eigenp}) with $E,j,\kappa$ as $p_{Ej\kappa,i}$. The mode solutions of Dirac equation with MIT boundary condition can be written as:
\begin{equation}
U_k^{\mathrm{MIT}}(x)=C_k^{\mathrm{MIT}}U_k(x),
\end{equation}
where $k$ represents:
\begin{equation}\label{spk2}
k=(E,j,m_j,\kappa,i).
\end{equation}
The normalization constant will be calculated in Sec. \ref{IIIB3}.

\subsubsection{Spectrum energy}

Now we show there is no $E\widetilde{E}<0$ modes, thus the rotating and nonrotating vacua are equivalent for MIT boundary condition case. First, we consider the case when $M=0$. The eigen equation (\ref{eigenp}) becomes:
\begin{equation}\label{mitp}
j_{j-\frac{1}{2}}(pR)=\pm j_{j+\frac{1}{2}}(pR),
\end{equation}
where the plus or minus sign depends on the signs of $\kappa$ and $\varsigma$.
The roots of Eq. (\ref{mitp}) times $R$ are the zeros of functions 
\begin{equation}
J_{j}(x)\pm J_{j+1}(x), 
\end{equation}
where $J_n$ is the $n$th Bessel function. According to the theorems in \cite{mathBessel2}, the first nonzero zero $\xi_{j,1}^{-}$ of $J_{j}(x)-J_{j+1}(x)$ satisfies $\xi_{j,1}^{-}>\xi'_{j,1}$, and the first nonzero zero $\xi_{j,1}^{+}$ of $J_{j}(x)+J_{j+1}(x)$ satisfies $\xi_{j,1}^{+}>\xi'_{j+1,1}$, where $\xi'_{j,1}$ is the first zero of $J'_{j}(x)$. Using the property \cite{mathBessel1}:
\begin{equation}
\xi'_{j,1}>\sqrt{j(j+2)},
\end{equation}
we get 
\begin{equation}
\xi_{j,1}^{\pm}>j\geq m_j.
\end{equation}
Thus 
\begin{equation}\label{pc}
|E_{jm_j\kappa,i}|R\geq p_{Ej\kappa,i}R > m_j.
\end{equation}
Combine Eq.~(\ref{pc}) with Eq.~(\ref{pr}) and we prove that there is no $E\widetilde{E}<0$ modes for $M=0$ case.

When $M \neq 0$, we assume $E>0$ ($E<0$ case can be proved in a similar way). Let us first consider the ordinary case ($\varsigma=1$), the eigen equation (\ref{eigenp}) becomes:
\begin{equation}
\begin{aligned}
&J_{j}(pR)-\frac{p}{E+M}J_{j+1}(pR)=0,\ \ \ \kappa>0,\\
&J_{j}(pR)+\frac{E+M}{p}J_{j+1}(pR)=0,\ \ \ \kappa<0.
\end{aligned}
\end{equation}

For the equation with $\kappa>0$, we set the first nonzero root $pR=\xi_{j,1}$. Since $0<\frac{p}{E+M}<1$, one has $J_j(\xi_{j,1})<J_{j+1}(\xi_{j,1})$. However, in the interval $0<x\leq\xi_{j,1}^{-}$, $J_{j}(x)>J_{j+1}(x)>0$, so $\xi_{j,1}>\xi_{j,1}^{-}>m_j$, thus Eq. (\ref{pc}) is satisfied.

For the equation with $\kappa<0$, use $\frac{p}{E+M}>0$ and $J_{j}(x)>J_{j+1}(x)>0$ when $0<x\leq \xi_{j,1}^{-}$, one can easily know the first nonzero root is larger than $\xi_{j,1}^{-}$, thus Eq. (\ref{pc}) is satisfied.

Now let's turn to the chiral case ($\varsigma=-1$), the eigen equations (\ref{eigenp}) becomes:
\begin{equation}
\begin{aligned}
&J_{j}(pR)+\frac{p}{E+M}J_{j+1}(pR)=0,\ \ \ \kappa>0,\\
&J_{j}(pR)-\frac{E+M}{p}J_{j+1}(pR)=0,\ \ \ \kappa<0.
\end{aligned}
\end{equation}

For the equation with $\kappa>0$, use $\frac{p}{E+M}>0$ and $J_{j}(x)>J_{j+1}(x)>0$ when $0<x\leq \xi_{j,1}^{-}$, one can easily know the first nonzero root is larger than $\xi_{j,1}^{-}$, thus Eq. (\ref{pc}) is satisfied.

For the equation with $\kappa>0$, it can be written in the following form:
\begin{equation}
p\frac{J_j(pR)}{J_{j+1}(pR)}=M+E,
\end{equation}
which is the formula (3.49) in Ref \cite{rcylinder}. Then one can follow the proof below (3.49) in \cite{rcylinder} and finally Eq. (\ref{pc}) is satisfied.  

Yet, we have proved there is no $E\widetilde{E}<0$ modes and thus the rotating and nonrotating vacua are equivalent for the MIT boundary condition case.
\subsubsection{Normalization}\label{IIIB3}
To perform the second quantization for MIT boundary condition case, we have to calculate the normalization constants $C_k^{\mathrm{MIT}}$. The inner product for two particle modes $U_k^{\mathrm{MIT}}$ and $U_{k'}^{\mathrm{MIT}}$ is:
\begin{equation}\label{MITen}
\begin{aligned}
\langle U_k^{\mathrm{MIT}},U_{k'}^{\mathrm{MIT}}\rangle &= C_k^{\mathrm{MIT}*}C_{k'}^{\mathrm{MIT}} \delta(k,k')[\mathfrak{I}_{j+\frac{1}{2}}^{+}(p_{Ej\kappa,i}R)\\&+\mathrm{sgn}(\kappa)\frac{M}{E}\mathfrak{I}_{j+\frac{1}{2}}^{-}(p_{Ej\kappa,i}R)],
\end{aligned}
\end{equation}
where $\delta(k,k')=\delta_{jj'}\delta_{m_j,m_j'}\delta_{\kappa,\kappa'}\delta_{ii'}\theta(EE')$, and $\mathfrak{I}_{j+\frac{1}{2}}^{\pm}$ are given by Eq. (\ref{inte}).

Combine the Eq. (\ref{eigenp}) with Eq. (\ref{MITen}), one can get the normalized constants:
\begin{equation}
C_k^{\mathrm{MIT}}=\frac{\sqrt{2}}{R|j_{j+\frac{1}{2}}(p_{Ej\kappa,i}R)|}\sqrt{\frac{E+M}{2ER-\varsigma(2j+1)+\varsigma \frac{M}{E}}},\ \ \kappa>0,
\end{equation}
\begin{equation}
C_k^{\mathrm{MIT}}=\frac{\sqrt{2}}{R|j_{j-\frac{1}{2}}(p_{Ej\kappa,i}R)|}\sqrt{\frac{E+M}{2ER+\varsigma(2j+1)+\varsigma \frac{M}{E}}},\ \ \kappa<0.
\end{equation}
The anti-particle modes are also normalized since the particle modes are normalized, and one can check $U_k^{\mathrm{MIT}}$ has the same normalization constant with its charge conjugate $V_k^{\mathrm{MIT}}$.

\subsubsection{Second quantization}

The second quantization can be performed as before. Expand the field by normalized modes:
\begin{equation}\psi_{\mathrm{MIT}}=\sum_{k} \theta\left(E\right)\left[U_{k}^{\mathrm{MIT}} \mathbf{b}_{k}^{\mathrm{MIT}}+V_{k}^{\mathrm{MIT}} \mathbf{d}_{k}^{\mathrm{MIT} \dagger}\right],\end{equation}
where
\begin{equation}\sum_{k} = \sum_{j=1/2}^{\infty}\  \sum_{m_{j}=-j}^{j}\ \  \sum_{\kappa=\pm({j}+ 1 / 2)}\ \sum_{i=1}^{\infty} \sum_{E=\pm\left|E\right|}.\end{equation}
The vacuum state $|0^{\mathrm{MIT}}\rangle$ for MIT boundary condition case is defined by 
\begin{equation}
\mathbf{b}_k^{\mathrm{MIT}}|0^{\mathrm{MIT}}\rangle=\mathbf{d}_k^{\mathrm{MIT}}|0^{\mathrm{MIT}}\rangle=0.
\end{equation}

Thus, we have finished the field quantization for the Dirac field in rotating coordinates with two kinds of boundary conditions.

\section{Fermion condensate}\label{IV}

In this section, we calculate the thermal expectation value of fermion condensate in an thermal equilibrium rigidly-rotating sphere. The spectral and MIT boundary conditions are considered separately. 

We calculate the fermion condensate $\overline{\psi}\psi$ in a straight forward way. The field operator $\psi(x)$ and $\overline{\psi}(x)$ can be expanded by creation and annihilation operators:
\begin{equation}
\begin{aligned}
&\psi(x)=\sum_k\theta(E_k)C_k[b_k U_k(x)+d_k^{\dagger}V_k(x)],\\
&\overline{\psi}(x)=\sum_k\theta(E_k)C_k^{*}[d_k \overline{V}_k(x)+b_k^{\dagger}\overline{U}_k(x)],
\end{aligned}
\end{equation}
where $C_k$ are the normalization constants. We have
\begin{equation}
\begin{aligned}
\langle\overline{\psi}\psi\rangle =\sum_{kk'}\theta(E_k)\theta(E_{k'})&C_k^{*}C_{k'}[\langle b_k^{\dagger}b_{k'}\rangle \overline{U}_kU_{k'}+\langle d_kd_{k'}^{\dagger}\rangle \overline{V}_kV_{k'}\\&+\langle b_k^{\dagger}d_{k'}^{\dagger}\rangle \overline{U}_kV_{k'}+\langle d_kb_{k'}\rangle \overline{V}_kU_{k'}],
\end{aligned}
\end{equation}
where $\langle \cdot \rangle$ means the ensemble average for a thermal equilibrium rotating system. According to \cite{Vilenkin,Landau5}, 
\begin{equation}
\begin{aligned}
&\langle b_k^{\dagger}b_{k'}\rangle=\frac{1}{e^{\beta(\widetilde{E_k}-\mu)}+1}\delta(k,k'),\\
&\langle d_kd_{k'}^{\dagger}\rangle=1-\langle d_{k'}^{\dagger} d_k\rangle
=(1-\frac{1}{e^{\beta(\widetilde{E_k}+\mu)}+1})\delta(k,k'),\\
&\langle b_k^{\dagger}d_{k'}^{\dagger}\rangle=\langle d_kb_{k'}\rangle =0.
\end{aligned}
\end{equation}
Using  $V_k(x)=i\gamma^{\hat{2}}U_k^{*}(x)$, one has $\overline{V}_kV_k=-\overline{U}_kU_k$. Let 
\begin{equation}
\begin{aligned}
w(\widetilde{E_k})&=(1-\langle d_{k}^{\dagger}d_k\rangle-\langle b_k^{\dagger}b_{k}\rangle)\theta(E_k)\\&=\frac{\theta(E_k)}{2}(\mathrm{tanh}\frac{\beta(\widetilde{E_k}-\mu)}{2}+\mathrm{tanh}\frac{\beta(\widetilde{E_k}+\mu)}{2}),
\end{aligned}
\end{equation}
then the condensate can be written as:
\begin{equation}
\langle\overline{\psi}\psi\rangle =-\sum_{k}{|C_k|}^2w(\widetilde{E_k})\overline{U}_kU_{k}.
\end{equation}
Set
\begin{equation}
\begin{aligned}
A_{jm_j\kappa i}(r,\theta)=\mathrm{sgn}(\kappa)\frac{1}{2}[&j_{j-\frac{1}{2}}^2(p_k r)(\chi_{jm_j}^{+})^\dagger \chi_{jm_j}^{+}\\&-j_{j+\frac{1}{2}}^2(p_k r)(\chi_{jm_j}^{-})^\dagger \chi_{jm_j}^{-}],
\end{aligned}
\end{equation}
and
\begin{equation}
\begin{aligned}
B_{jm_j\kappa i}(r,\theta)=\frac{M}{2E}[&j_{j-\frac{1}{2}}^2(p_k r)(\chi_{jm_j}^{+})^\dagger \chi_{jm_j}^{+}\\&+j_{j+\frac{1}{2}}^2(p_k r)(\chi_{jm_j}^{-})^\dagger \chi_{jm_j}^{-}].
\end{aligned}
\end{equation}
We have
\begin{equation}
\overline{U_k}U_k=A_{jm_j\kappa i}+B_{jm_j\kappa i}.
\end{equation}
Finally the condensate can be expressed as:

\begin{equation}\label{conden}
\begin{aligned}
\langle\overline{\psi}\psi\rangle=-\sum_{j=1/2}^{\infty}\ &\sum_{\kappa=\pm}\sum_{i=1}^{\infty}\ \sum_{m_j=-j}^{j}\\&{|C_{jm_j\kappa i}|}^2w(\widetilde{E}_{jm_j\kappa i})(A_{jm_j\kappa i}+B_{jm_j\kappa i}).
\end{aligned}
\end{equation}

\subsection{spectral boundary condition}
For the case of spectral boundary condition, one notices that $p_{j,m_j,\kappa, i}=p_{j, -m_j, -\kappa, i}$, $(\chi_{j,m_j}^{\pm})^\dagger\chi_{j,m_j}^{\pm}=(\chi_{j,-m_j}^{\pm})^\dagger\chi_{j,-m_j}$, $C_{j,m_j,\kappa,i}=C_{j,-m_j,-\kappa,i}$, so one has
\begin{equation}
A_{j,m_j,\kappa,i}=-A_{j,-m_j,-\kappa,i},\ \ \ \ B_{j,m_j,\kappa,i}=B_{j,-m_j,-\kappa,i}.
\end{equation}
Thus the expression of the condensate can be simplified as:
\begin{widetext}
\begin{equation}\label{spc}
\begin{aligned}
\langle\overline{\psi}\psi\rangle =-\sum_{j=1/2}^{\infty}\  \sum_{\kappa=\pm}\ \sum_{i=1}^{\infty}{|C_{j\kappa i}|}^2\sum_{m_j=1/2}^{j}\ \{
[w(\widetilde{E}_{jm_j\kappa i})-w(\overline{E}_{jm_j\kappa i})]A_{jm_j\kappa i}+[w(\widetilde{E}_{jm_j\kappa i})+w(\overline{E}_{jm_j\kappa i})]B_{jm_j\kappa i}\},
\end{aligned}
\end{equation}
\end{widetext}
where $\overline{E}=E+\Omega m_j$ and $C_{j\kappa i}$ is the abbreviation of $C_{j\kappa i,m_j>0}$, which only depends on $j,\kappa, i$ when $m_j>0$.
The condensate is a function of $\theta$ and $r$ in general. In the special case $\Omega=0$, i.e., nonrotating case, one can simplify the expression further. Using the additional formula:
\begin{equation}
\begin{aligned}
\sum_{m}Y_{lm}(\theta,\phi)Y_{lm}^*(\theta',\phi')&=\sum_{m}Y_{lm}^*(\theta,\phi)Y_{lm}(\theta',\phi')\\&=\frac{2l+1}{4\pi}P_l(\mathrm{cos}\Theta),
\end{aligned}
\end{equation}
where $\mathrm{cos}\Theta=\mathrm{cos}\theta\mathrm{cos}\theta'+\mathrm{sin}\theta\mathrm{sin}\theta'\mathrm{cos}(\phi-\phi')$, we can get
\begin{equation}
\sum_{m_j=-j}^{j}(\chi_{jm_j}^{\pm})^\dagger\chi_{jm_j}^{\pm}=\frac{2j+1}{4\pi}P_{j\mp\frac{1}{2}}(\cos{\Theta}=1)=\frac{2j+1}{4\pi}.
\end{equation}
Then Eq. (\ref{spc}) with $\Omega=0$ can be simplified as:
\begin{equation}
\begin{aligned}
\langle\overline{\psi}\psi\rangle =-\sum_{j=1/2}^{\infty}\  \sum_{\kappa=\pm}\ \sum_{i=1}^{\infty}{|C_{j\kappa i}|}^2  w({E}_{j\kappa i})\mathfrak{B}_{j\kappa i,m>0},
\end{aligned}
\end{equation}
where $E_{j\kappa i}$ is the abbreviation of $E_{j\kappa i,m_j>0}$ and 
\begin{equation}
\begin{aligned}
\mathfrak{B}_{j\kappa i,m>0}&=\sum_{m_j=1/2}^{j}2B_{jm_j\kappa i}(r,\theta)\\&=\frac{M}{2E}\frac{2j+1}{4\pi}[j_{j-\frac{1}{2}}^2(p_{j\kappa i,m>0}r)+j_{j+\frac{1}{2}}^2(p_{j\kappa i,m>0}r)],
\end{aligned}
\end{equation}
which only depends on $r$. So the condensate inside a nonrotating sphere with spectral boundary condition only depends on coordinate $r$. This is easy to understand since a nonrotating spherical system has spherical symmetry.

The condensate we calculate above is divergent. We can subtract its divergent part which is independent of temperature to get the condensate $\langle :\overline{\psi}\psi:\rangle$ which is finite. To do this, we just need to replace $\omega(\widetilde{E})$ by 
\begin{equation}
\omega'(\widetilde{E})=-\frac{\theta(E)}{1+e^{\beta(\widetilde{E}-\mu)}}-\frac{\theta({E})}{1+e^{\beta(\widetilde{E}+\mu)}}.
\end{equation}

\begin{figure*}[htbp]
\centering

\subfigure[ $\ M=1,\ \beta=2,\ \theta=\pi/2$ ]{
\begin{minipage}{0.45\textwidth}
\centering
\includegraphics[width=1\linewidth]{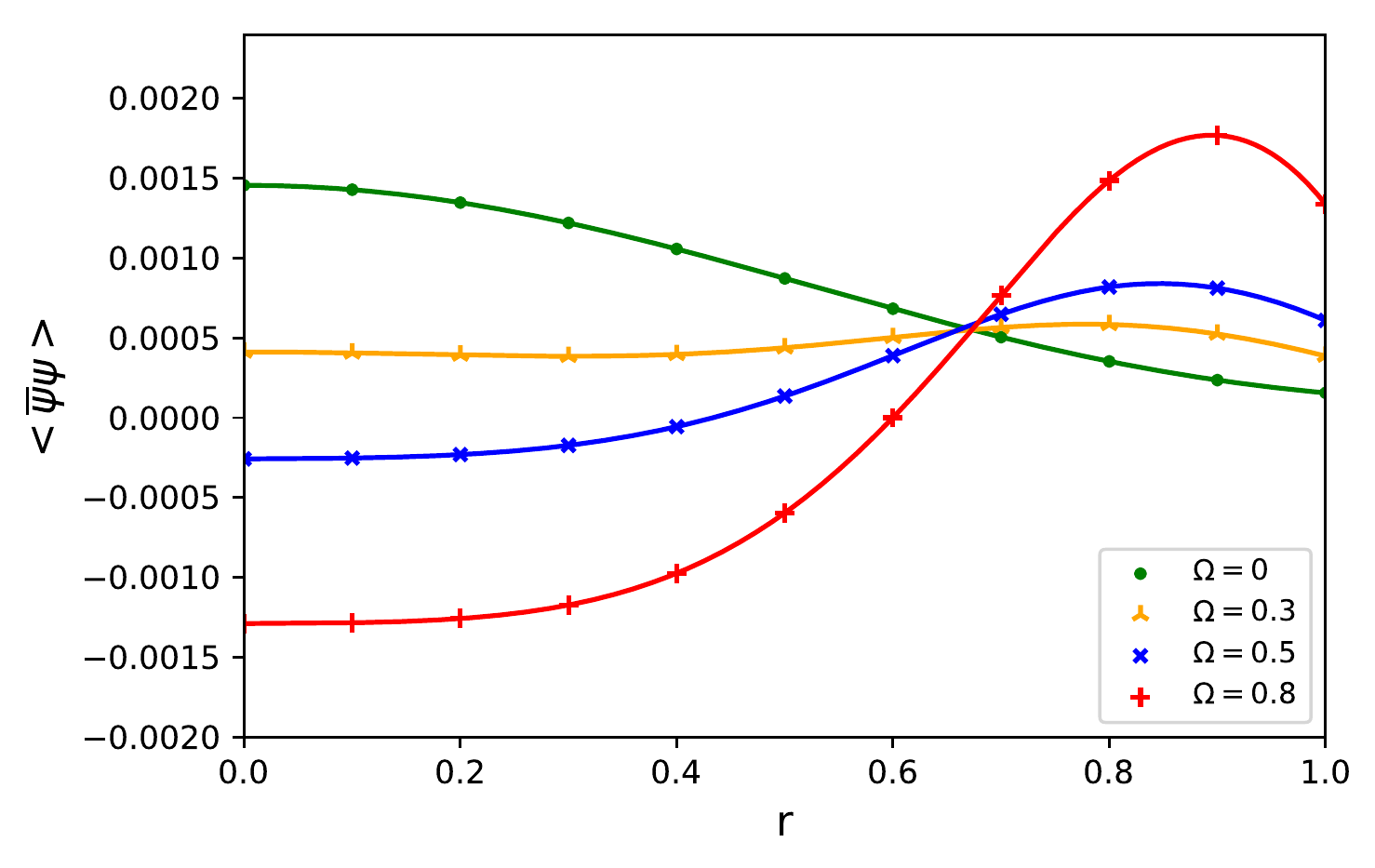} 
\label{fig1a}
\end{minipage}
}
\subfigure[$\ M=1,\ \beta=0.5,\ \theta=\pi/2$]{
\begin{minipage}{0.42\textwidth}
\centering
\includegraphics[width=1\linewidth]{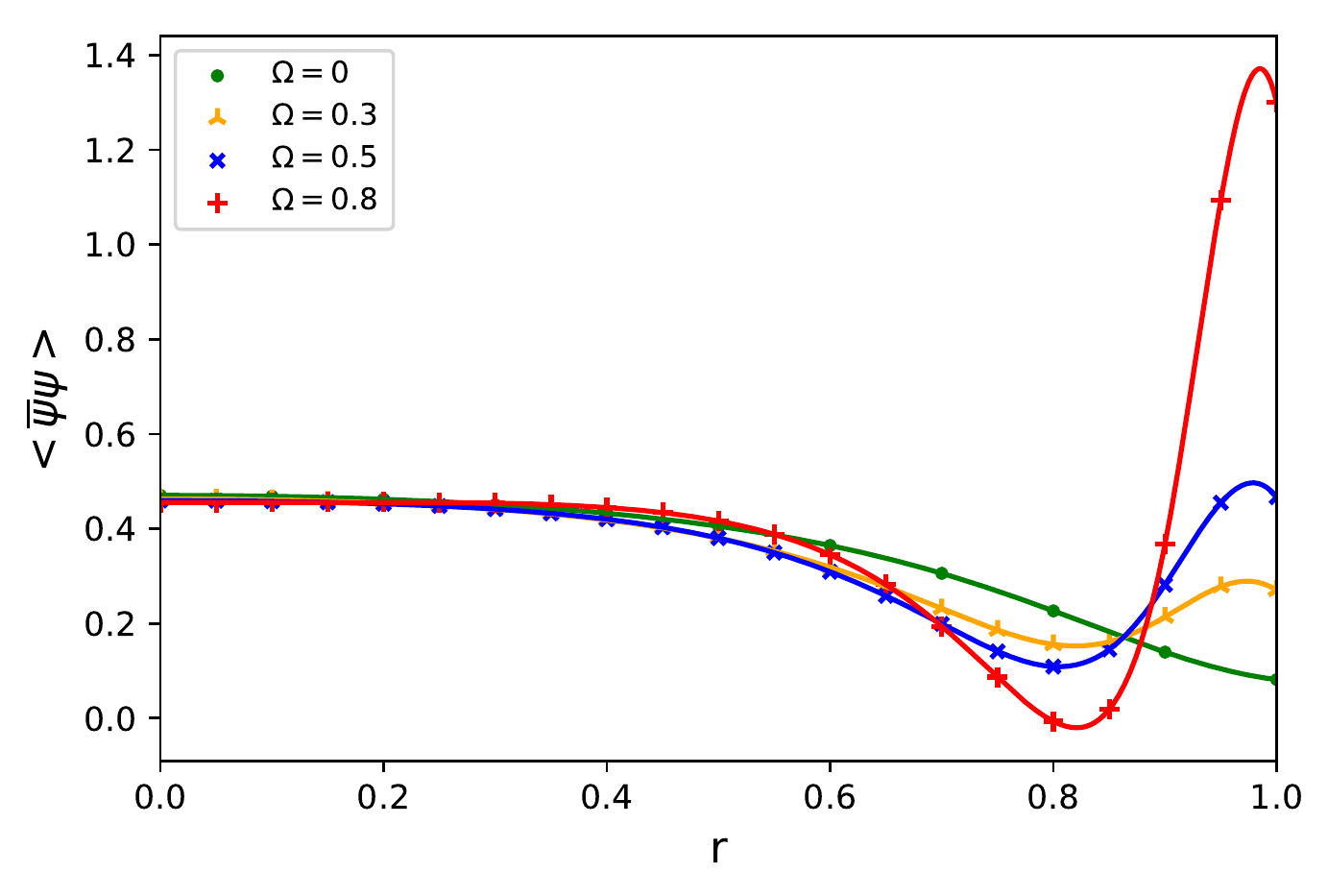} 
\label{fig1b}
\end{minipage}
}
\subfigure[ $\ M=1,\ \Omega=0.5,\ \theta=\pi/2$ ]{
\begin{minipage}[t]{0.43\textwidth}
\centering
\includegraphics[width=1\linewidth]{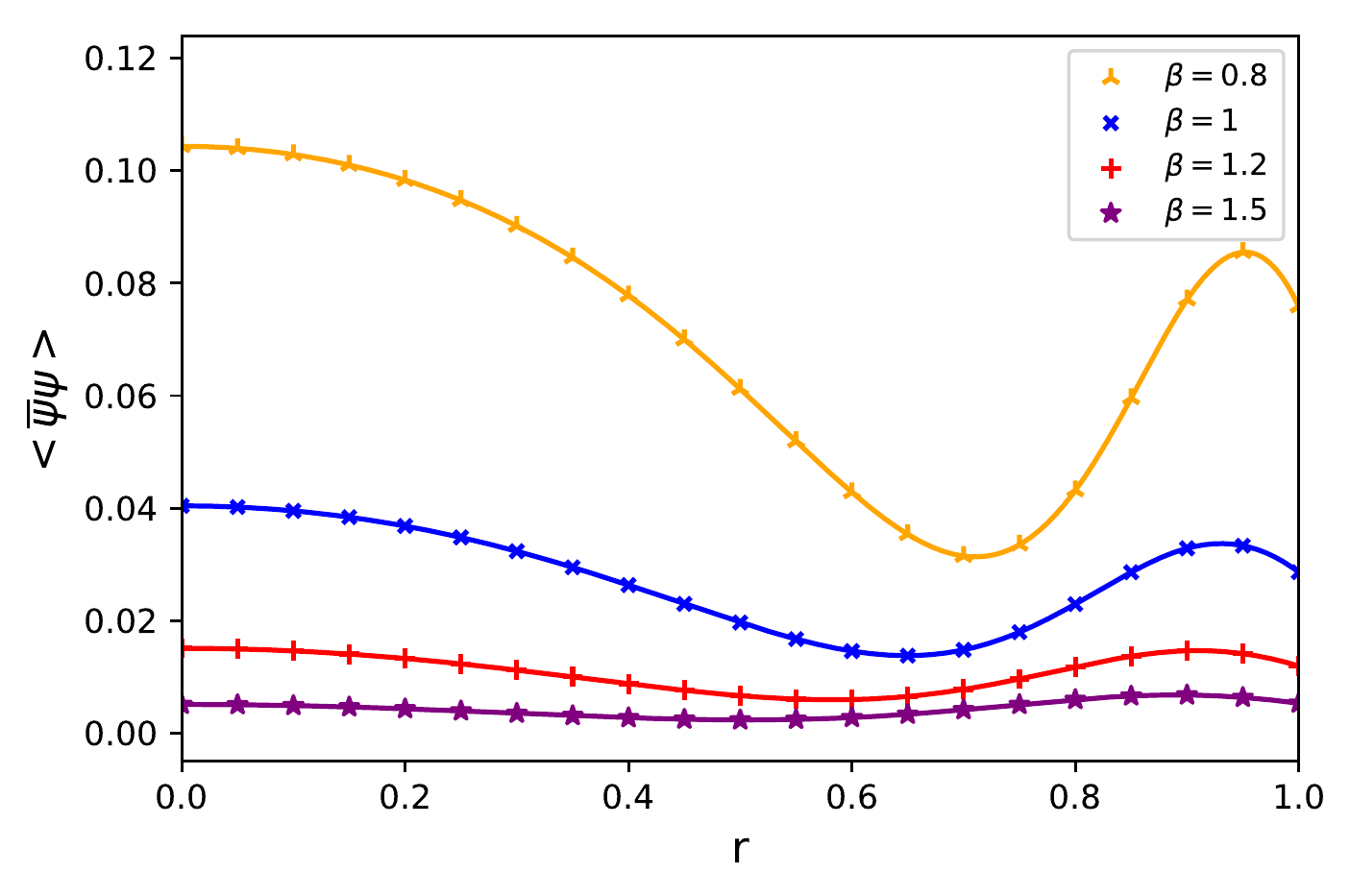} 
\end{minipage}
\label{fig1c}
}
\subfigure[ $\ \beta=1,\ \Omega=0.5,\ \theta=\pi/2$ ]{
\begin{minipage}[t]{0.44\textwidth}
\centering
\includegraphics[width=1\linewidth]{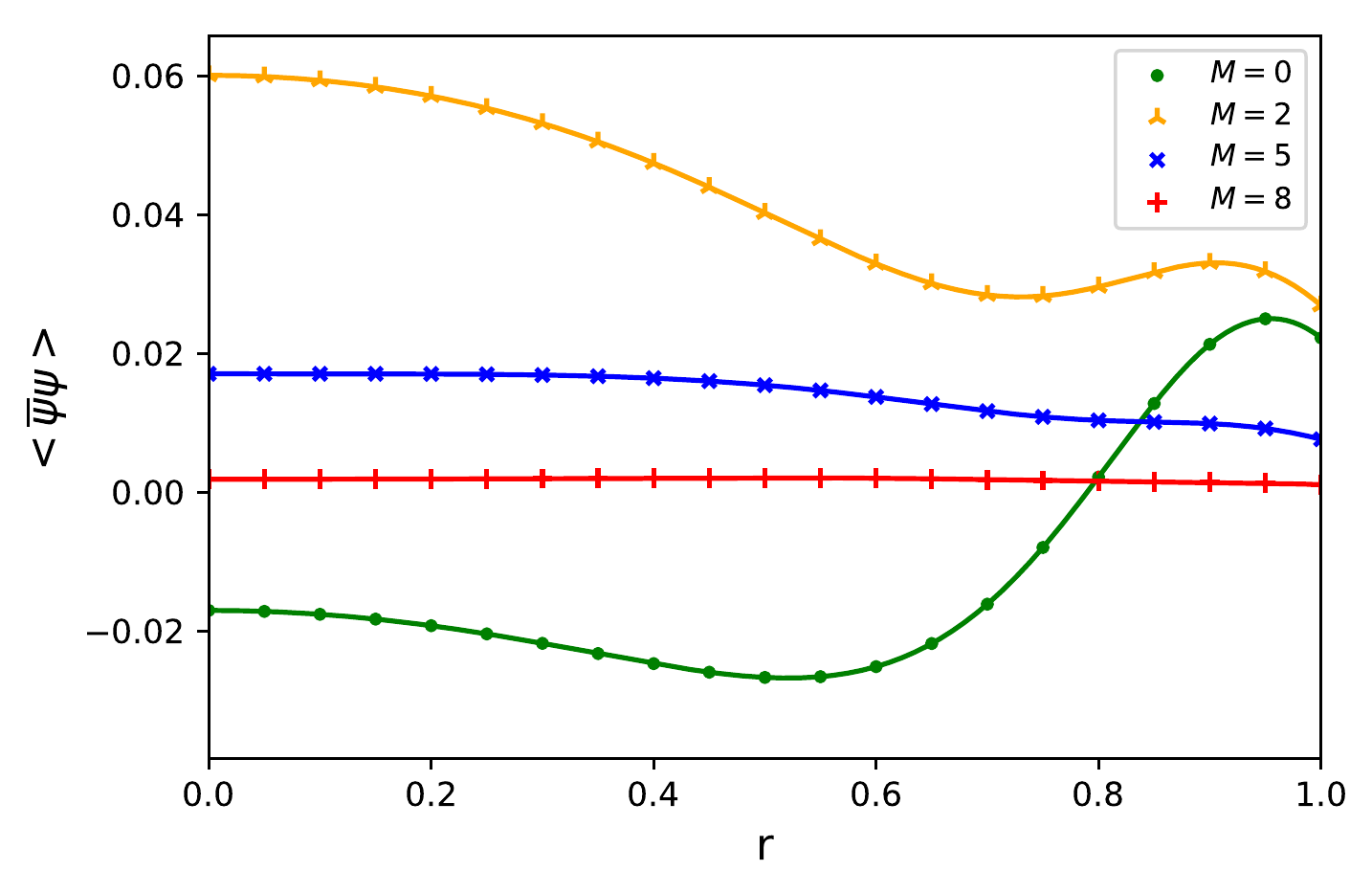} 
\label{fig1d}
\end{minipage}
}
\subfigure[ $\ M=1,\ \beta=2,\ \Omega=0.8$ ]{
\begin{minipage}[t]{0.45\textwidth}
\centering
\includegraphics[width=1\linewidth]{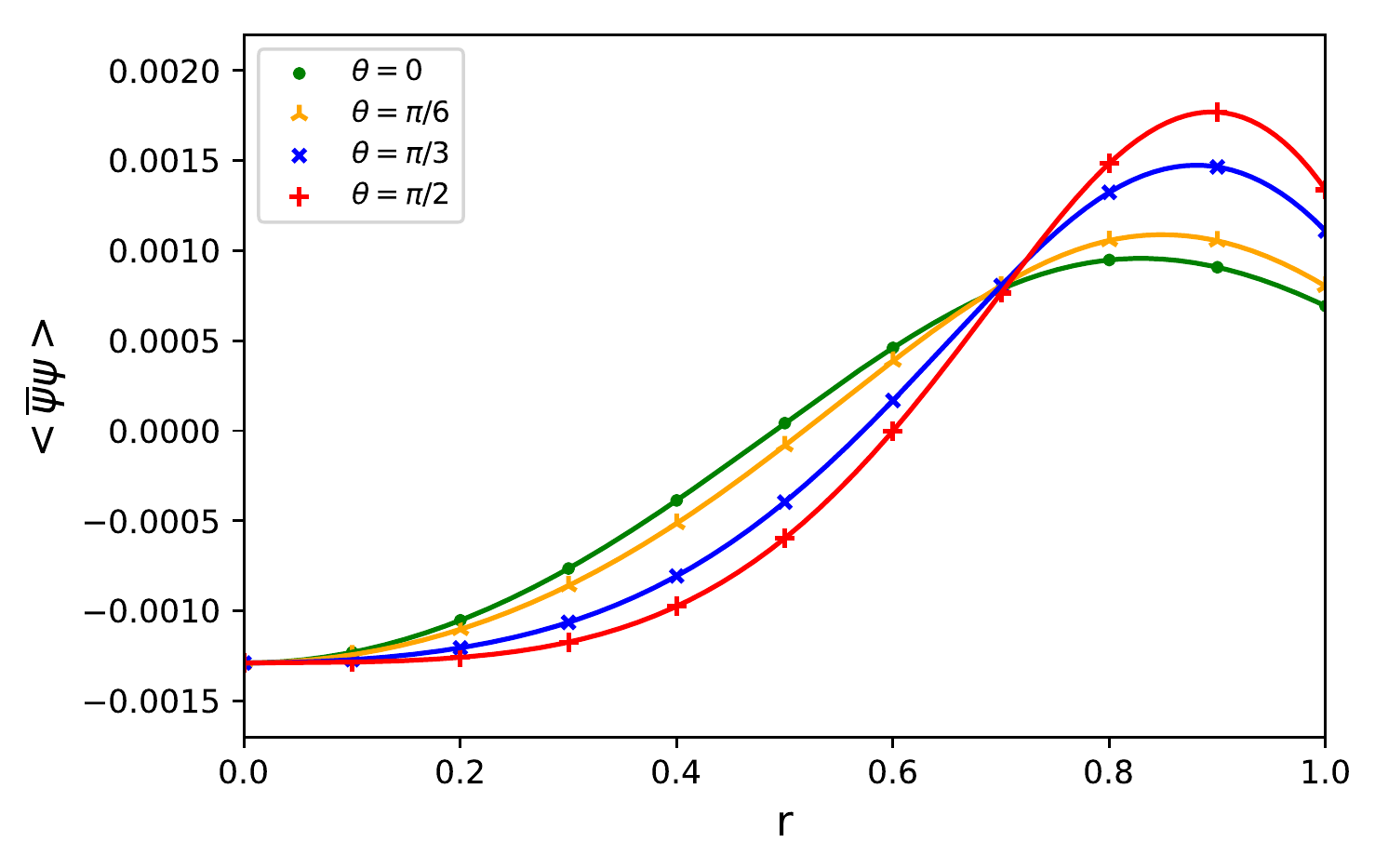} 
\label{fig1e}
\end{minipage}
}
\subfigure[ $\ M=1,\ \beta=0.5,\ \Omega=0.8$ ]{
\begin{minipage}[t]{0.42\textwidth}
\centering
\includegraphics[width=1\linewidth]{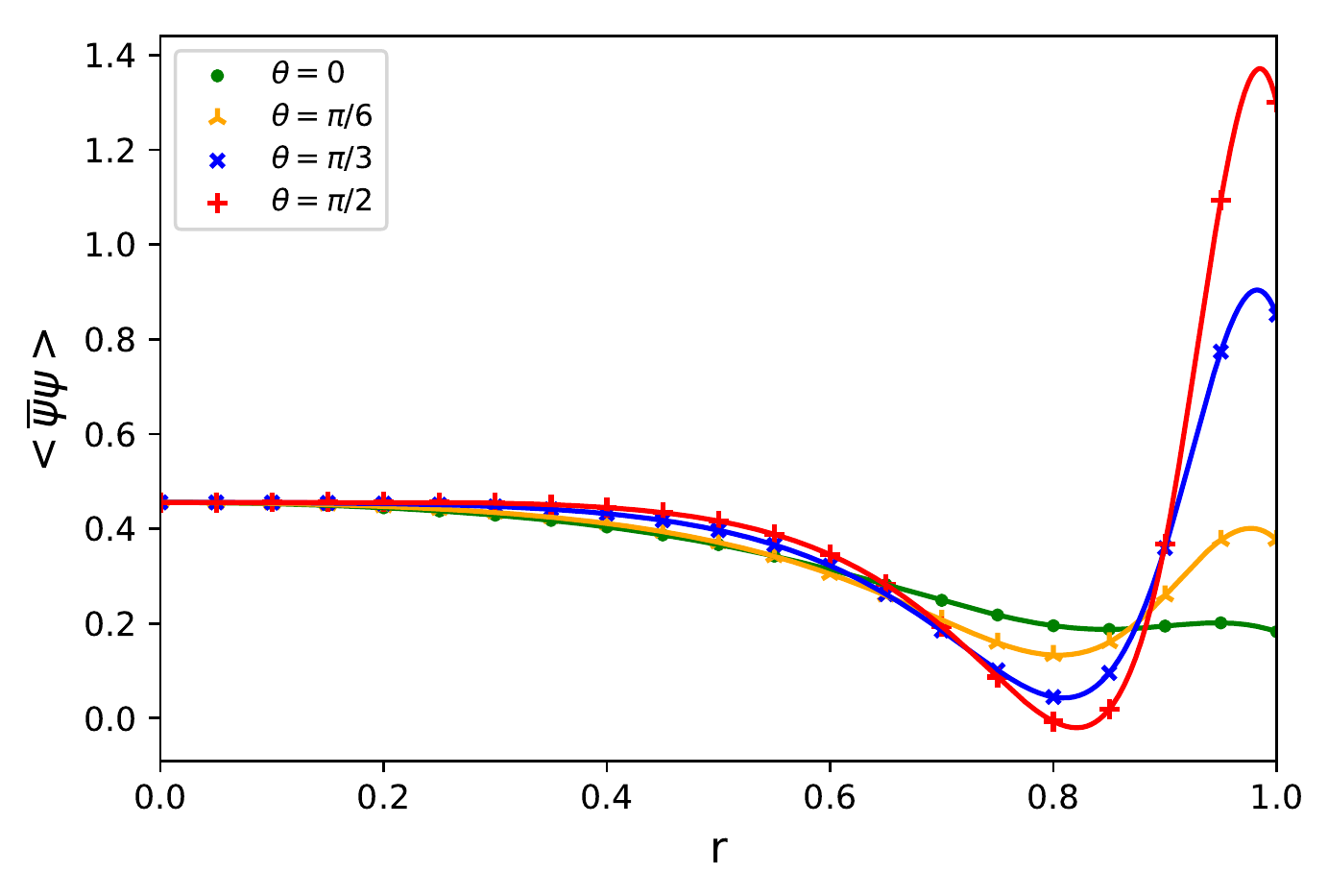} 
\label{fig1f}
\end{minipage}
}
\centering
\caption{The thermal expectation values of fermion condensate $\langle:\overline{\psi}\psi:\rangle$ for spectral boundary condition. The radius of the sphere is set to be $R=1$. All results are calculated at chemical potential $\mu=0$. (a) The condensate as a function of $r$ at $\theta=\pi/2$ at $M=1$, fixed inverse temperature $\beta=2$ and various angular velocity $\Omega$. (b) The condensate as a function of $r$ at $\theta=\pi/2$ at  $M=1$, fixed inverse temperature $\beta=0.5$ and various angular velocity $\Omega$. (c)  The condensate as a function of $r$ at $\theta=\pi/2$ at $M=1$, fixed angular velocity $\Omega=0.5$ and various inverse temperature $\beta$. (d) The condensate as a function of $r$ at $\theta=\pi/2$ at fixed inverse temperature $\beta=1$, angular velocity $\Omega=0.5$ and various mass $M$. (e) The condensate as a function of $r$ at several $\theta$ angles at $M=1$, fixed inverse temperature $\beta=2$ and angular velocity $\Omega=0.8$. (f) The condensate as a function of $r$ at several $\theta$ angles at $M=1$, fixed inverse temperature $\beta=0.5$ and angular velocity $\Omega=0.8$.}
\label{fig1}
\end{figure*}

In Fig. \ref{fig1}, we present some numerical results of the fermion condensate $\langle:\overline{\psi}\psi:\rangle$ for the spectral boundary condition. Fig. \ref{fig1a} and Fig. \ref{fig1b} show the influence of the rotation on the fermion condensate. We can observe that rotation increases the condensate at large $r$. And this effect becomes stronger when the rotation speed increases. Fig. \ref{fig1c} shows the influence of the inverse temperature $\beta$ on the fermion condensate. The fermion condensate increases when the temperature increases, which is consistent with the result in \cite{rcylinder}. The influence of the mass $M$ on the fermion condensate is presented in Fig. \ref{fig1d}. Fig. \ref{fig1e} and Fig. \ref{fig1f} show that the effects of rotation are different at different $\theta$ angles. The rotation has stronger effect when $\theta$ is close to $\pi/2$. For the spectral boundary condition case, the fermion condensate on the boundary is finite and nonzero in general. There are some differences between our results of fermion condensate and that in \cite{rcylinder} for cylindrical spectral boundary condition. The fermion condensate in \cite{rcylinder} is zero when the particle is massless, which is not the case here. Another difference is that the fermion condensate in \cite{rcylinder} is always positive while it can be negetive here.  
\subsection{MIT boundary condition}
\begin{figure*}[htbp]
\centering

\subfigure[ $\ M=1,\ \beta=2,\ \theta=\pi/2$ ]{
\begin{minipage}{0.45\textwidth}
\centering
\includegraphics[width=1\linewidth]{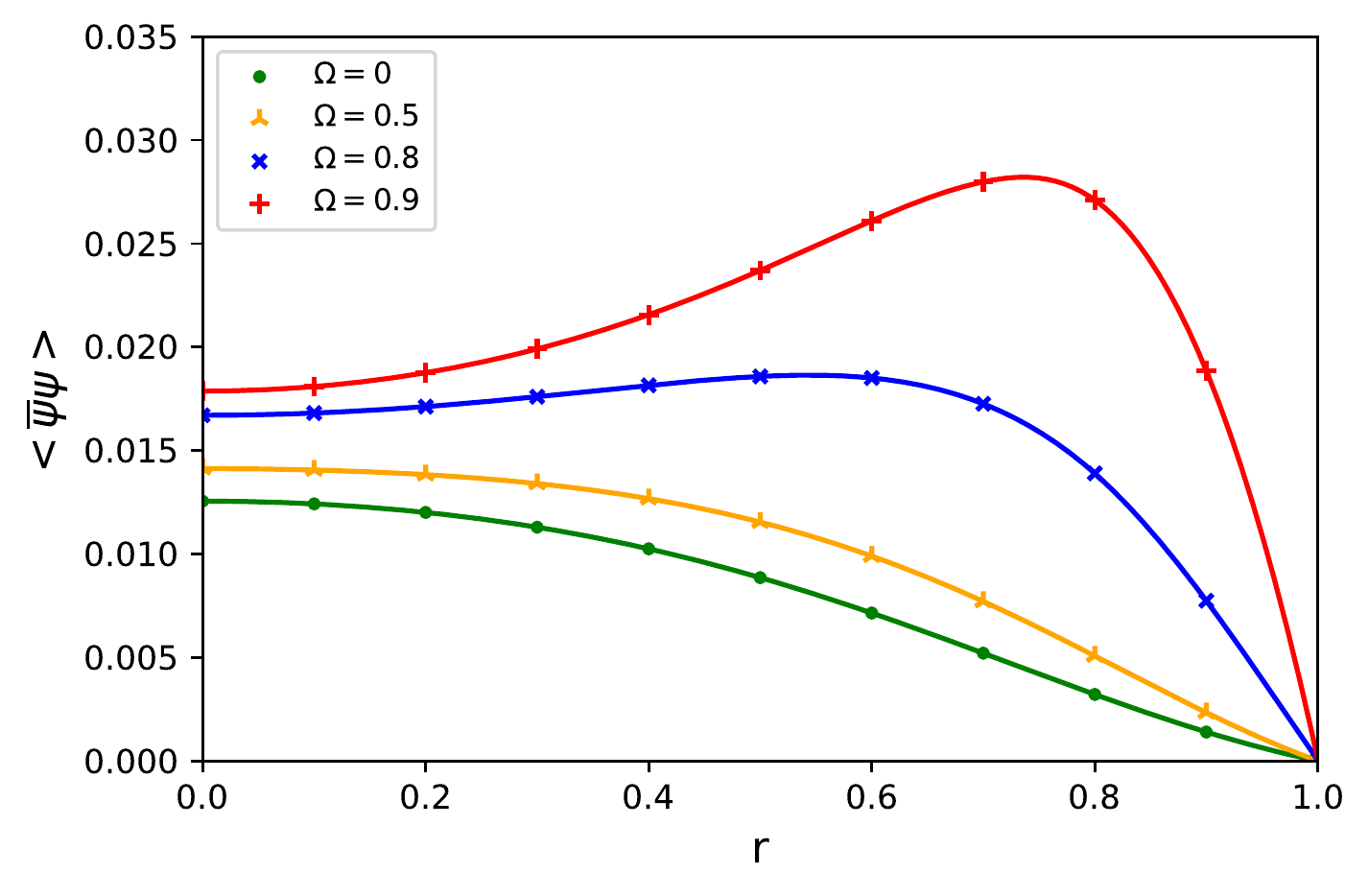} 
\label{fig2a}
\end{minipage}
}
\subfigure[$\ M=1,\ \beta=0.5,\ \theta=\pi/2$]{
\begin{minipage}{0.42\textwidth}
\centering
\includegraphics[width=1\linewidth]{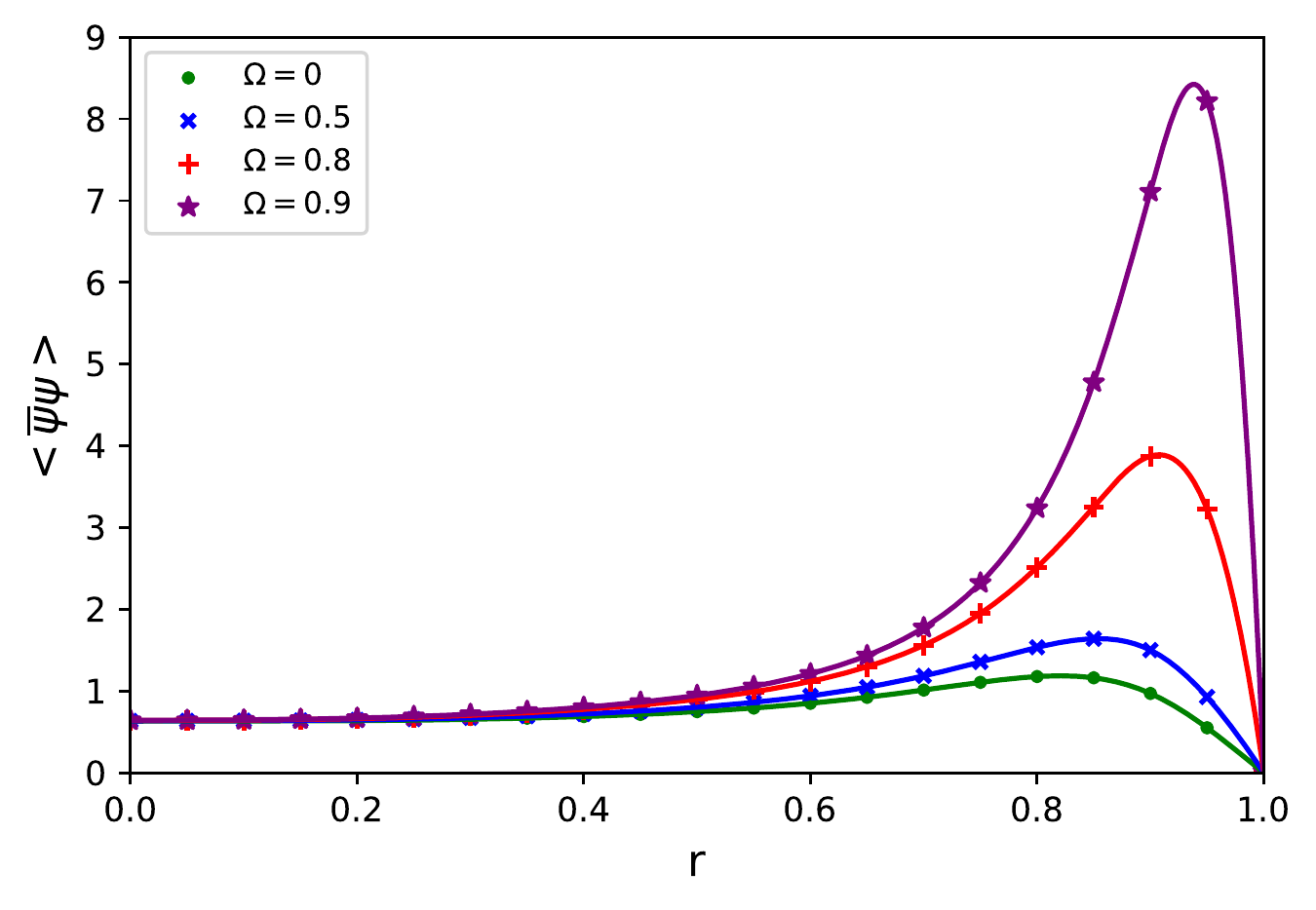} 
\label{fig2b}
\end{minipage}
}
\subfigure[ $\ M=1,\ \Omega=0.5,\ \theta=\pi/2$ ]{
\begin{minipage}[t]{0.43\textwidth}
\centering
\includegraphics[width=1\linewidth]{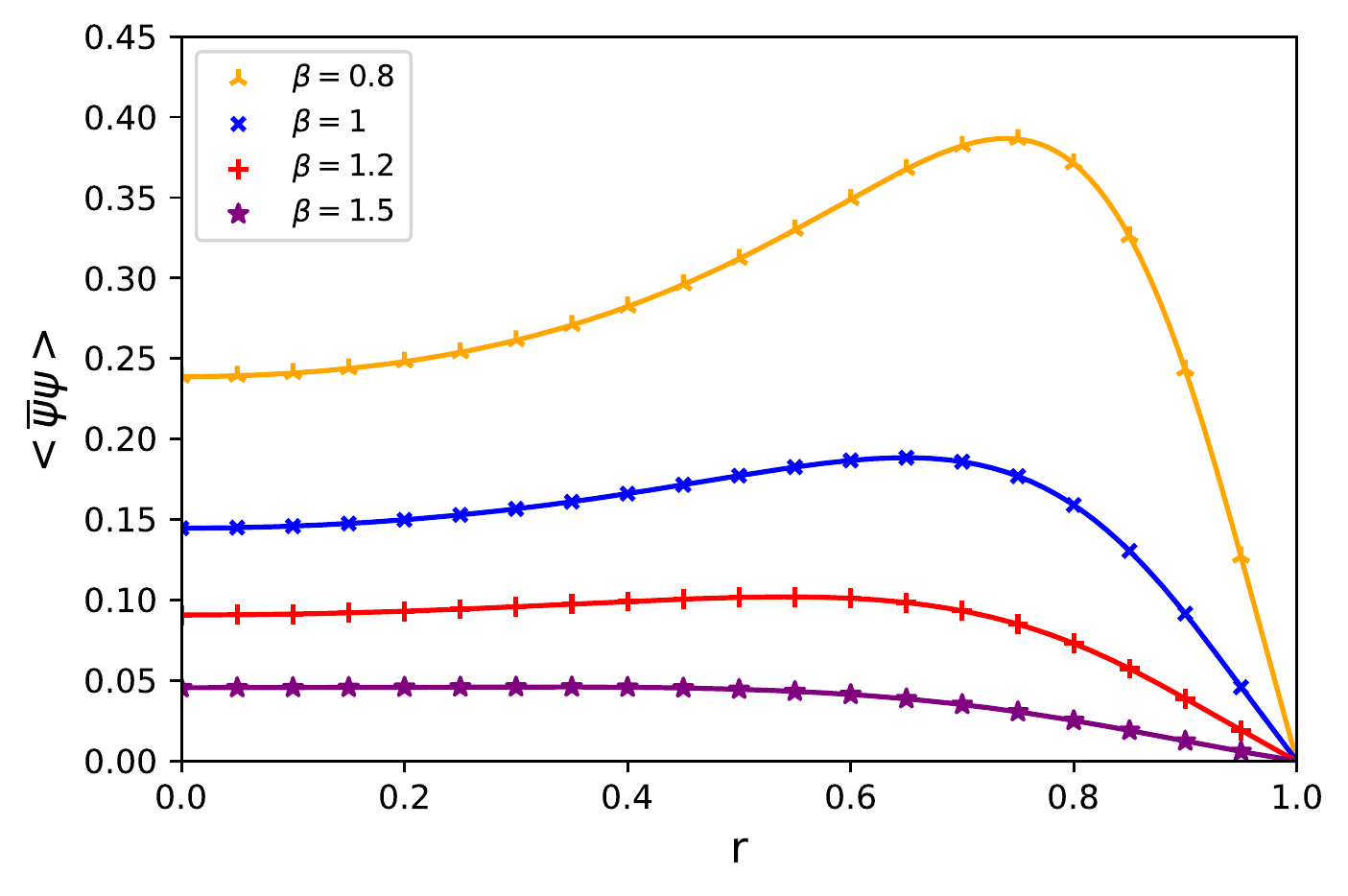} 
\end{minipage}
\label{fig2c}
}
\subfigure[ $\ \beta=1,\ \Omega=0.5,\ \theta=\pi/2$ ]{
\begin{minipage}[t]{0.43\textwidth}
\centering
\includegraphics[width=1\linewidth]{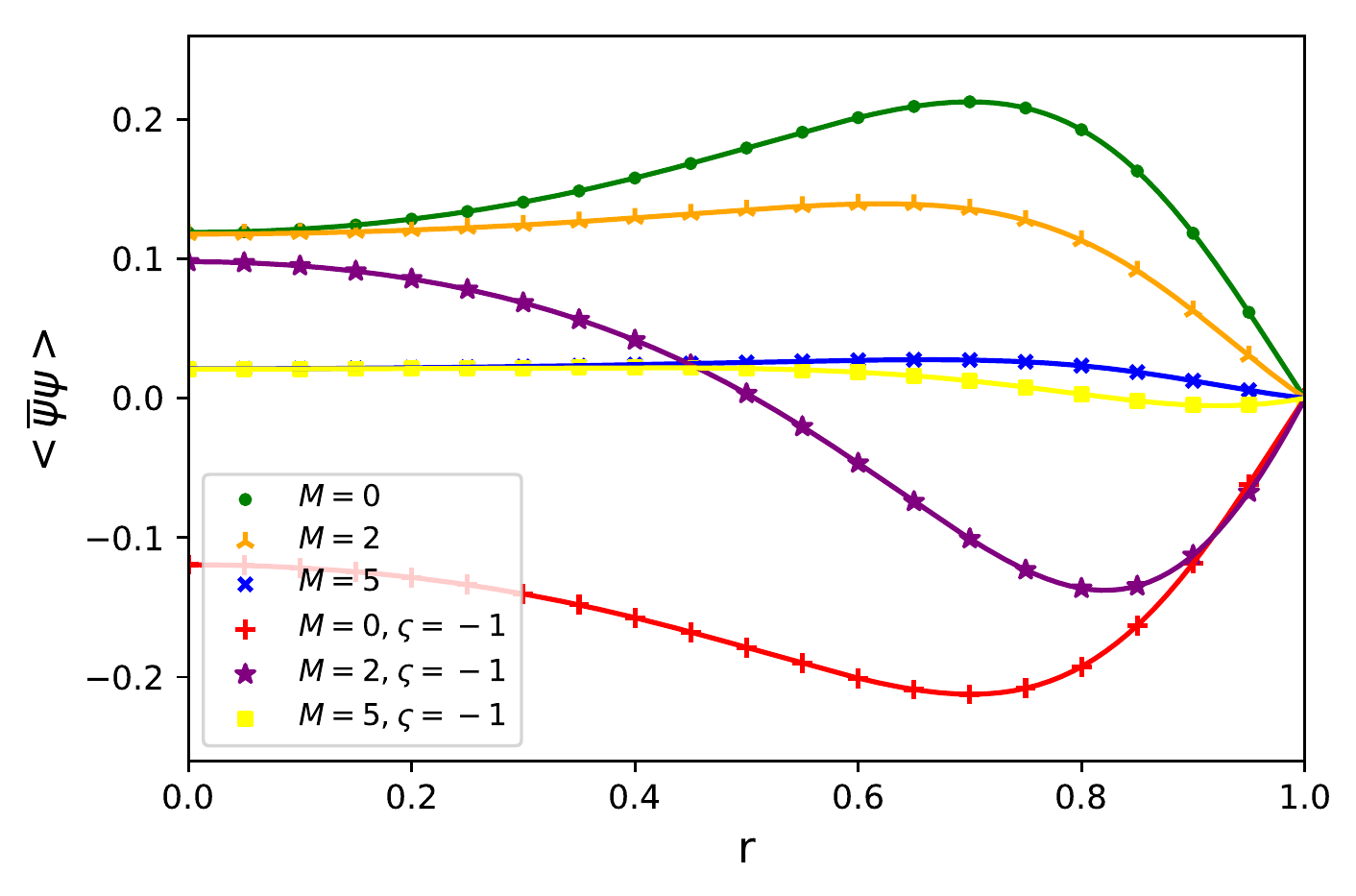} 
\label{fig2d}
\end{minipage}
}
\subfigure[ $\ M=1,\ \beta=2,\ \Omega=0.8$ ]{
\begin{minipage}[t]{0.45\textwidth}
\centering
\includegraphics[width=1\linewidth]{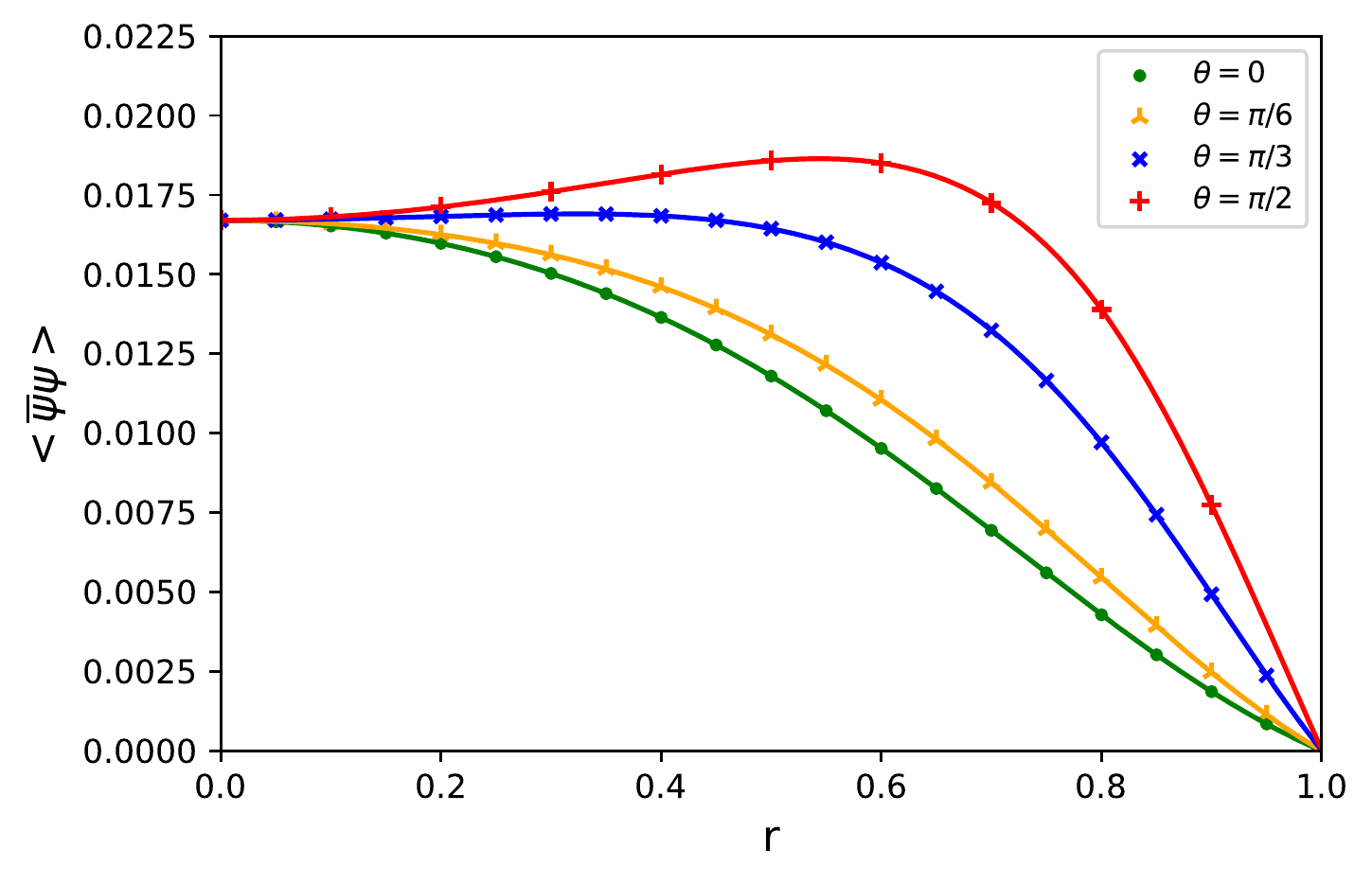} 
\label{fig2e}
\end{minipage}
}
\subfigure[ $\ M=1,\ \beta=0.5,\ \Omega=0.8$ ]{
\begin{minipage}[t]{0.43\textwidth}
\centering
\includegraphics[width=1\linewidth]{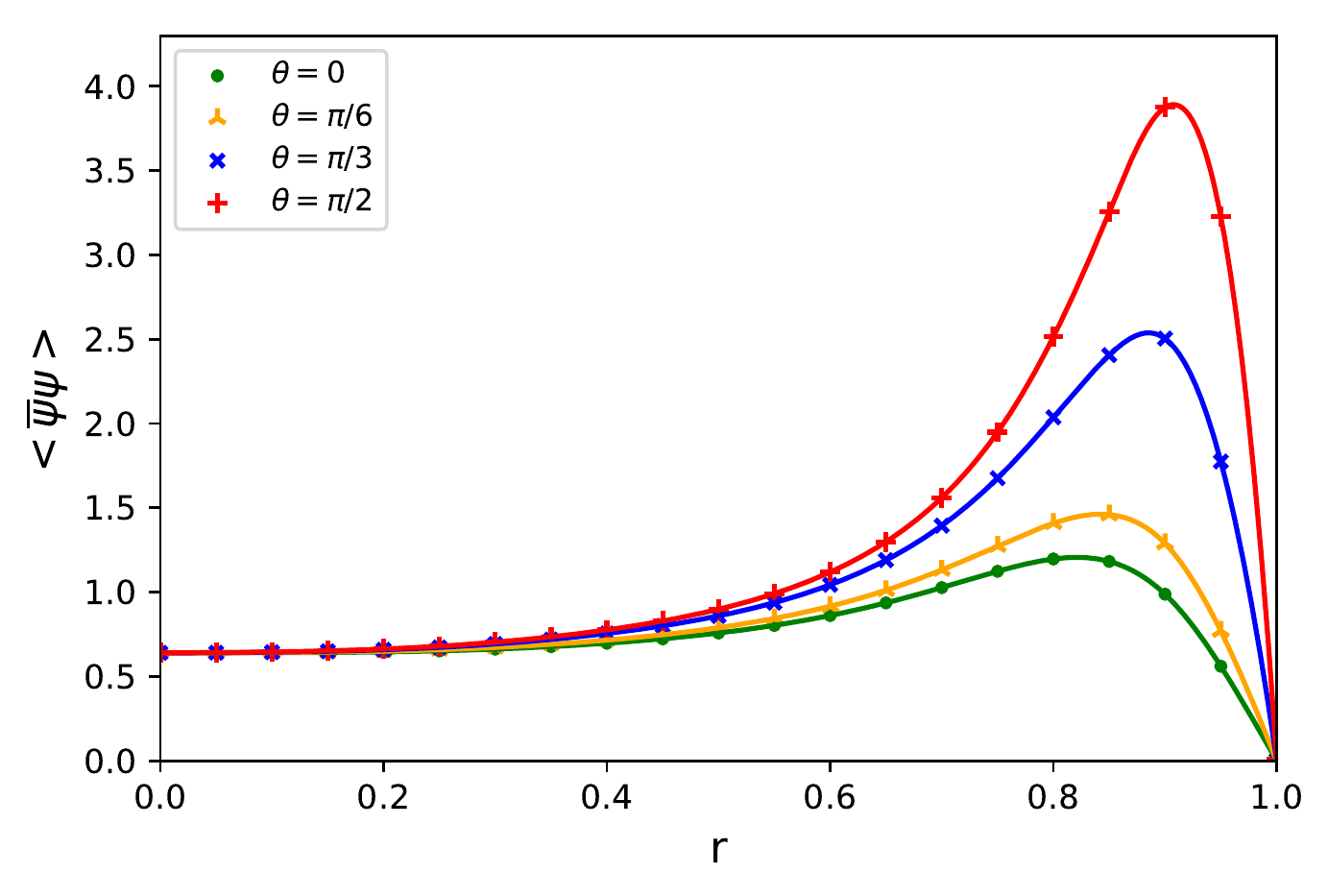} 
\label{fig2f}
\end{minipage}
}
\centering
\caption{The thermal expectation values of fermion condensate $\langle:\overline{\psi}\psi:\rangle$ for MIT boundary condition. The radius of the sphere is set to be $R=1$. All results are calculated at chemical potential $\mu=0$. We take $\varsigma=1$ if not noted.  (a) The condensate as a function of $r$ at $\theta=\pi/2$ at $M=1$, fixed inverse temperature $\beta=2$ and various angular velocity $\Omega$. (b) The condensate as a function of $r$ at $\theta=\pi/2$ at $M=1$, fixed inverse temperature $\beta=0.5$ and various angular velocity $\Omega$. (c) The condensate as a function of $r$ at $\theta=\pi/2$ at $M=1$, fixed angular velocity $\Omega=0.5$ and various inverse temperature $\beta$. (d) The condensate as a function of $r$ at $\theta=\pi/2$ with $\varsigma=\pm1$ at fixed inverse temperature $\beta=1$, angular velocity $\Omega=0.5$ and various mass $M$. (e) The condensate as a function of $r$ at several $\theta$ angles at $M=1$, fixed inverse temperature $\beta=2$ and angular velocity $\Omega=0.8$. (f) The condensate as a function of $r$ at several $\theta$ angles at $M=1$, fixed inverse temperature $\beta=0.5$ and angular velocity $\Omega=0.8$. }
\label{fig2}
\end{figure*}
For MIT boundary condition, the discretized momentum $p_{Ej\kappa i}$ and normalization constants $C_{j\kappa i}$ are independent of $m_j$. Consider $(\chi_{j,m_j}^{\pm})^\dagger\chi_{j,m_j}^{\pm}=(\chi_{j,-m_j}^{\pm})^\dagger\chi_{j,-m_j}$, we have 
\begin{equation}
A_{j,m_j,\kappa,i}=A_{j,-m_j,\kappa,i},\ \ \ \ B_{j,m_j,\kappa,i}=B_{j,-m_j,\kappa,i}.
\end{equation}
Thus the condensate Eq. (\ref{conden}) can be simplified as:
\begin{equation}
\begin{aligned}
\langle\overline{\psi}\psi\rangle=&-\sum_{j=1/2}^{\infty} \sum_{\kappa=\pm}\sum_{i=1}^{\infty}{|C_{j\kappa i}|}^2\sum_{m_j=1/2}^{j}\\&[ w(\widetilde{E}_{jm_j\kappa i})+w(\overline{E}_{jm_j\kappa i})](A_{jm_j\kappa i}+B_{jm_j\kappa i}),
\end{aligned}
\end{equation}
which is a function of $\theta$ and $r$. When the system is nonrotating, i.e., $\Omega=0$, we can also simplify the expression further:
\begin{equation}
\langle\overline{\psi}\psi\rangle =-\sum_{j=1/2}^{\infty}\ \sum_{\kappa=\pm}\sum_{i=1}^{\infty}{|C_{j\kappa i}|}^2w(E_{j\kappa i})(\mathfrak{A}_{j\kappa i}+\mathfrak{B}_{j\kappa i}),
\end{equation}
where
\begin{equation}
\begin{aligned}
\mathfrak{A}_{j\kappa i}&=\sum_{m_j=-j}^{j}A_{jm_j\kappa i}(r,\theta)\\&=\mathrm{sgn}(\kappa)\frac{2j+1}{4\pi}\frac{1}{2}[j_{j-\frac{1}{2}}^2(p_k r)-j_{j+\frac{1}{2}}^2(p_k r)],\\
\mathfrak{B}_{j\kappa i}&=\sum_{m_j=-j}^{j}B_{jm_j\kappa i}(r,\theta)\\&=\frac{M}{2E}\frac{2j+1}{4\pi}[j_{j-\frac{1}{2}}^2(p_k r)+j_{j+\frac{1}{2}}^2(p_k r)].
\end{aligned}
\end{equation}
So the condensate inside a nonrotating sphere with MIT boundary condition also only depends on coordinate $r$. 

In Fig .\ref{fig2}, we present some numerical results of the fermion condensate $\langle:\overline{\psi}\psi:\rangle$ for the MIT boundary condition case. One noticeable feature is that the condensate vanishes on the boundary, which is different from the spectral boundary condition case. This feature was also noticed in \cite{rcylinder}. We can show this feature does not depend on the shape of the boundary, because it is a direct consequence of the MIT boundary condition:
\begin{equation}
\begin{aligned}
\overline{\psi}\psi|_{x_b}&=(-i\psi^\dagger\gamma^{i\dagger} n_i\gamma^{0})(i\gamma^{i\dagger} n_i\psi)|_{x_b}\\&=\overline{\psi}(\gamma^{i}n_i)^2\psi|_{x_b}=-\overline{\psi}\psi|_{x_b}=0,
\end{aligned}
\end{equation}
where we used Eq. (\ref{cg}) and $n_\mu=(0,-n_i)$.
Fig. \ref{fig2a} and Fig. \ref{fig2b} show the influence of the rotation on the condensate. Like the spectral boundary condition case, the rotation increases the condensate at large $r$. The influences of the inverse temperature $\beta$ and mass $M$ are presented in Fig. \ref{fig2c} and Fig. \ref{fig2d} respectively. Like the spectral boundary condition case, the fermion condensate increases when the temperature increases. Fig. \ref{fig2e} and Fig. \ref{fig2f} show that the effects of rotation are stronger at $\theta$ angle which is closer to $\pi/2$. This is also similar with the spectral boundary condition case. But, the quantitative results with MIT boundary condition are different from that with the spectral boundary condition.

Yet, we have finished the calculation of the condensate inside a sphere with spectral and MIT boundary conditions. We find that the expectation value of the condensate depends on the boundary condition.

\section{Summary}\label{V}
In this paper, we studied a Dirac field enclosed inside a sphere in Minkowski space-time. The solutions to the Dirac equation in rotating spherical coordinates have the same form with the solutions to the Dirac equation in nonrotating spherical coordinates, but the spectrum in rotating spherical coordinates is different from that in nonrotating spherical coordinates by a term relative to the rotation speed. To constrain the system inside the speed of light surface, we considered two kinds of boundary conditions, namely, the spectral and MIT boundary conditions. The rotating quantum states of the system inside the sphere have been constructed for each boundary condition. And the equivalence of the rotating and nonrotating vacua when the boundary is placed inside the speed of light surface was proved. Combining our proof and the proof for cylindrical boundary case in \cite{rcylinder}, one expect that any possible physical field in rotating coordinates has a unique quantization scheme, and the rotating vaccum is identical to the Minkowski vacuum. Thus, a rotating observer will not see the Minkowski vacuum has strange effects like Unruh effect.

Finally, the thermal expectation value of the fermion condensate for a thermal equilibrium rotating field was calculated. We found it depends on the boundary condition, and of course, varies with coordinates and rotation speed. Calculating thermal expectation values has some practical meaning for the study of rapidly rotating matter. For example, in noncentral high energy heavy-ion collisions (HICs), the strong interacting matter can carry large angular momentum and reach very high angular velocity \cite{vorticalfluid}. The great vorticity can cause some interesting effects, one is the so called chiral vortical effect \cite{chiralv1,chiralv2,chiralv3}, which is analogy to the chiral magnetic effect \cite{FukushimaCHM,Yinke,CMEreview}. This effect predicts a non-vanishing current along the rotating axis, whose thermal expectation can be calculated based on the spectrum of fermions in the rotating frame \cite{Vilenkin,rcylinder}. Great vorticity can also influence the phase transition of the matter \cite{YinJiang}. In HICs, it is shown that the phase conversion from
the hadron phase to the quark phase can be influenced by rotation \cite{zzhang2}. To characterize this phase conversion, the fermion condensate can be used as an order parameter \cite{NJLreview}. However, the quark-gluon systems created in HICs, also called as the "fireballs", are estimated to be of only 2-10 fm in diameter. \cite{Palhares2011}. The finite size effects can influence the phase transition as well \cite{reviewK}. To consider both the rotation and finite size effects for the phase transition in HICs, the fermion condensate calculated in this paper may have possible applications, which we will study in the future. 
\section*{Acknowledgements}
This work is supported in part by the National Natural Science Foundation of China (under Grants No. 11475085, No. 11535005, No. 11905104, and No. 11690030) and by Nation Major State Basic Research and Development of China (2016YFE0129300). X. Luo is supported by the National Key Research and Development Program of China (2018YFE0205201),  the National Natural Science Foundation of China (Grants No. 11828501, No. 11575069, No. 11890711 and No. 11861131009).

\bibliography{ref}
\end{document}